\documentclass[a4paper,fleqn,usenatbib]{mnras}

\usepackage[T1]{fontenc}
\usepackage{ae,aecompl}

\usepackage{graphicx}	     % Including figure files
\usepackage{amsmath}	% Advanced maths commands
\usepackage{amssymb}	% Extra maths symbols

\usepackage{amsmath}
\def\thet{{\tilde{\theta}}}
\def\phit{{\tilde{\phi}}}
\def\thetp{{\tilde{\theta}^\prime}}
\def\phitp{{\tilde{\phi}^\prime}}
\def\d{\mathrm{d}}
\title[Emission from the cocoon of a GRB jet]{Thermal and non-thermal emission from the cocoon of a gamma-ray burst jet}

\author[Fabio De Colle et al.]{
Fabio De Colle,$^{1}$\thanks{E-mail: fabio@nucleares.unam.mx}
Wenbin Lu,$^{2}$
Pawan Kumar,$^{2}$
Enrico Ramirez-Ruiz,$^{3}$
George Smoot$^{4,5,6}$
\\
$^{1}$Instituto de Ciencias Nucleares, Universidad Nacional Aut\'onoma de M\'exico, A. P. 70-543 04510 D. F. Mexico\\
$^{2}$Department of Astronomy, University of Texas at Austin, Austin, TX 78712, USA\\
$^{3}$Department of Astronomy and Astrophysics, University of California, Santa Cruz, California 95064\\
$^{4}$Helmut and Ana Pao Sohmen Professor at Large, Institute for Advanced Study, Hong Kong University of Science and Technology,\\ Clear Water Bay, Kowloon, Hong Kong\\
$^{5}$PCCP; APC, Universit\' e Paris Diderot, Universit\'e Sorbonne Paris Cit\'e,  75013 France\\
$^{6}$BCCP; LBNL \& Physics Dept. University of California at Berkeley  CA 94720, USA
}

\date{Accepted XXX. Received YYY; in original form ZZZ}

\pubyear{2017}

\begin{document}
\label{firstpage}
\pagerange{\pageref{firstpage}--\pageref{lastpage}}
\maketitle

\begin{abstract}
We present hydrodynamic simulations of the hot cocoon produced when a relativistic jet passes through the gamma-ray burst (GRB) progenitor star and its environment, and we compute the lightcurve and spectrum of the radiation emitted by the cocoon. The radiation from the cocoon has a nearly thermal spectrum with a peak in the X-ray band, and it lasts for a few minutes in the observer frame; the cocoon radiation starts at roughly the same time as when $\gamma$-rays from a burst trigger detectors aboard GRB satellites. The isotropic cocoon luminosity ($\sim 10^{47}$~erg s$^{-1}$) is a few times smaller then the X-ray luminosity of a typical long-GRB afterglow during the plateau phase. This radiation should be identifiable in the Swift data because of its nearly thermal spectrum which is distinct from the somewhat brighter power-law component. The detection of this thermal component would provide information regarding the size and density stratification of the GRB progenitor star. Photons from the cocoon are also inverse-Compton (IC) scattered by electrons in a delayed jet. We present the IC lightcurve and spectrum, by post-processing the results of the numerical simulations. The IC spectrum lies in 10 keV--MeV band for typical GRB parameters. The detection of this IC component would provide an independent measurement of GRB jet Lorentz factor and it would also help to determine the jet magnetisation parameter.
\end{abstract}

\begin{keywords}
Hydrodynamics -- radiation mechanisms: non-thermal -- radiation mechanisms: thermal -- relativistic processes -- methods: numerical -- gamma-ray burst: general
\end{keywords}

%%%%%%%%%%%%%%%%%%%%%%%%%%%%%%%%%%%%%%%%%%%%%%%%%%

%%%%%%%%%%%%%%%%% BODY OF PAPER %%%%%%%%%%%%%%%%%%

%%%%%%%%%%%%%%%%%%%%%%%%%%%%%%%%%%%%%%%%%%%%%%%%%%%%%%%%%%%%%%%%%%%%%%%%%%%%%%
% INTRODUCTION
%%%%%%%%%%%%%%%%%%%%%%%%%%%%%%%%%%%%%%%%%%%%%%%%%%%%%%%%%%%%%%%%%%%%%%%%%%%%%%

\section{Introduction}

Long duration gamma-ray bursts (GRBs) are produced when the core of a
massive star collapses to a neutron star or a black hole
\citep[for recent reviews on GRBs see, e.g., ][]{piran04, woosley06a, fox06, gehrels09, kumar15}.
The newly formed compact object produces a pair of relativistic jets that
make their way out of the progenitor star along the polar regions. Punching their
way to the stellar surface, these jets shock heat
the material they encounter pushing it both sideways and along the jet's
direction. Therefore, the jet is surrounded by a hot cocoon made by this shock heated
plasma, which contributes to the collimation of the jet \citep{ramirez-ruiz02}.
The central engine activity is known to be highly variable and long-lived, giving multiple episodes of gamma-ray emission during the prompt phase \citep[e.g.,][]{ramirez-ruiz01b} and sometimes sharp increases in X-ray flux (flares) at much later time from minutes to hours \citep[e.g.,][]{chincarini07}.

The total amount of energy deposited in the cocoon (which is equal to the
work done by the jet on the medium it passes through) depends on the size
of the star and the jet luminosity. The temperature of the cocoon is
determined by the density of the star which controls the
expansion speed of the cocoon transverse
to the jet axis and hence the cocoon's volume. The temperature, energy,
and Lorentz factor of the cocoon are the main parameters that affect
its luminosity. Thus, the observation of radiation from the cocoon,
once it breaks out of the star, expands and later decelerates into the stellar
environment, should in principle provide information about the progenitor star.

The numerous GRB simulations performed to date \citep[e.g.,][]{zhang03,
mizuta06, morsony07, bromberg11, lazzati13, lopezcamara13, mizuta13, duffell15, bromberg16}
have considered the hydrodynamic or magnetohydrodynamic interactions
of the jet with the progenitor star, but not the radiation escaping the cocoon or the interaction of this
radiation with the relativistic jet.
However, it is the radiation from
the cocoon that provides information on the GRB progenitor star properties, and
that is a big part of the motivation for this work.

The only paper we know presenting cocoon lightcurves based on hydrodynamical simulation is by \citet{suzuki13}. However, this paper was mainly concerned with
the impact of the circumstellar medium on the early thermal X-ray emission.
Recently, \citet{nakar16} provided an analytic calculation of the 
cocoon radiation which includes the mixing of the
shocked jet and shocked stellar material, but not the IC interaction between
the cocoon and the jet. 

In this paper, we present lightcurves and spectra of radiation escaping
the cocoon by post-processing special relativistic hydrodynamic simulations
which follow the evolution of the GRB jet and the associated cocoon from
$\sim 10^8$ cm to $\sim 3 \times 10^{14}$ cm. The aforementioned mixing and its effect on the emergent cocoon radiation is build into our hydrodynamical simulations.
We also compute the radiation
from the cocoon scattered by electrons in the relativistic jet, which
produces higher energy photons. The non-thermal radio emission from the cocoon is studied in detail in a follow-up paper \citep{decolle18}.

The paper is organised as follows: in Section \ref{sec:methods} we provide
information about the pre-collapse Wolf-Rayet star, the numerical method
employed, and the simulation set up. The jet dynamics and the hydrodynamic
properties of the cocoon are discussed in Section \ref{sec:dynamics}. Details
of the calculation of the thermodynamic parameters and the radiation are
described in Section \ref{sec:cocoon}, where we also present lightcurves
and spectra for a number of selected jet and progenitor stellar
models. Section \ref{sec:ic} describes the inverse-Compton interaction
between the relativistic jet and the cocoon photons and the lightcurve of
the emergent high-energy photons; the main results of the paper are
discussed in Section \ref{sec:conclusions}. Throughout this paper, we use
the convention $G_{x} = G/10^{x}$ in cgs units.

%%%%%%%%%%%%%%%%%%%%%%%%%%%%%%%%%%%%%%%%%%%%%%%%%%%%%%%%%%%%%%%%%%%%%%%%%%%%%%
% METHODS
%%%%%%%%%%%%%%%%%%%%%%%%%%%%%%%%%%%%%%%%%%%%%%%%%%%%%%%%%%%%%%%%%%%%%%%%%%%%%%

\section{Methods and Initial conditions}
\label{sec:methods}

We run two-dimensional axisymmetric simulations using the adaptive mesh
refinement code \emph{Mezcal} \citep{decolle06, decolle12a,decolle12b,decolle12c},
which solves the special relativistic, hydrodynamics equations
on an adaptive grid. The simulations span six orders of magnitude in space
and time, and we consider the jet crossing the star and then moving
in the wind of the progenitor star.

%%%%%%%%%%%%%%%%%%%%%%%%%%%%%%%%%%%%
\begin{figure}
\centering
\includegraphics[width=8cm]{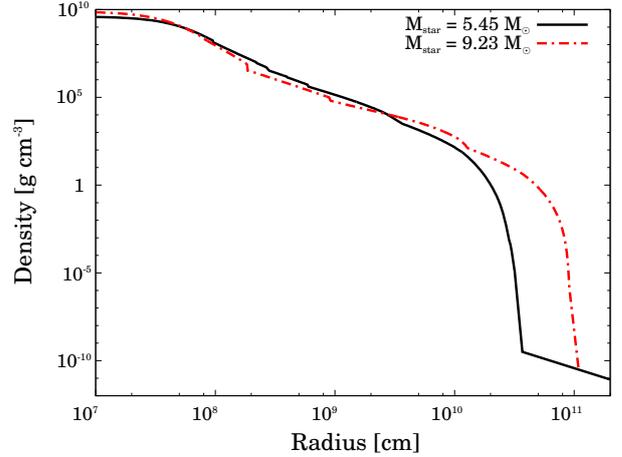}
\caption{Density stratification of the stellar pre-supernova models used as initial conditions in the numerical simulations. The two models are a 5.45 M$_\odot$ and a 9.23 M$_\odot$ pre-supernova stars (the E25 and 12TH models from \citealt{heger00} and \citealt{woosley06b} respectively).}
\label{fig1}
\end{figure}
%%%%%%%%%%%%%%%%%%%%%%%%%%%%%%%%%%%%

As progenitors of GRBs, we employ two pre-supernova, Wolf-Rayet stellar models (see Figure \ref{fig1}). The E25 stellar model \citep{heger00} is a star with an initial mass of $25 M_\odot$ reduced to a final mass of $5.45 M_\odot$ after losing its hydrogen and helium envelopes by massive winds.
The 12TH model \citep{woosley06b} has an initial mass of $12 M_\odot$, a final mass of $9.23 M_\odot$ and a more extended stellar envelope.
At a radius larger than the stellar radii of these models
($r > 3 \times 10^{10}$~cm and $r > 10^{11}$~cm for the E25 and 12TH models respectively), the ambient medium density is taken to
be that of the wind of the progenitor star which we assume had a
mass loss rate of $\dot{M}_w = 10^{-5}$ M$_\odot$ yr$^{-1}$ and a velocity of
$v_w = 10^8$~km~s$^{-1}$.

%%%%%%%%%%%%%%%%%%%%%%%%%%%%%%%%%%%%
\begin{figure*}
\centering
\includegraphics[width=\textwidth, angle = 0]{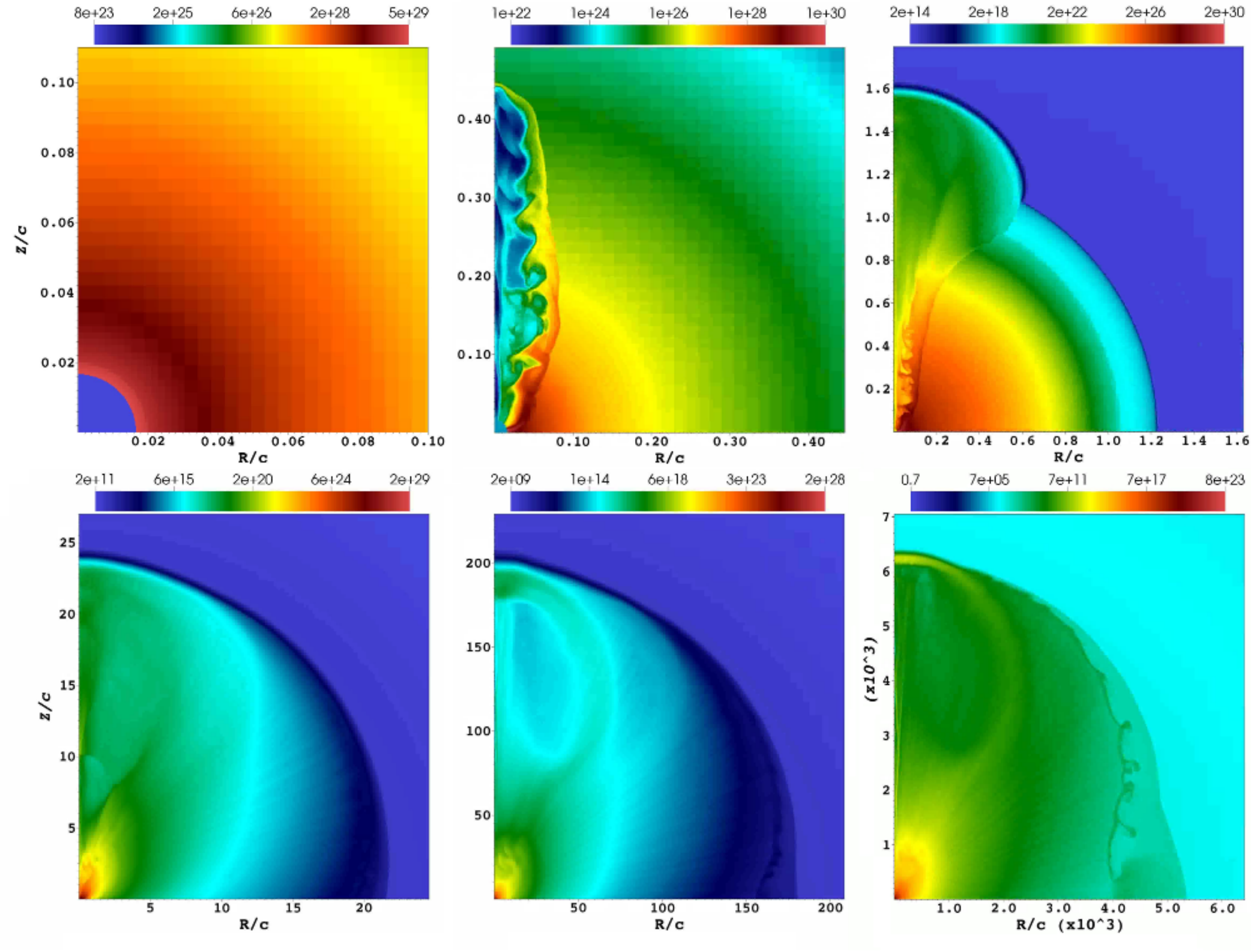}
\caption{Density map at different evolutionary times from the simulation with $\Gamma_{\rm j}=10$ and E25 progenitor. The plots correspond (top, left to right) to $t=0$ s, $t=5$ s, $t=6.4$ s
and (bottom, left to right) $t=28$ s, $t=200$ s, $t=6280$ s. In the third frame (at $t=6.4$ seconds) one can see that the cocoon has burst out of the progenitor star and is evolving in a less constrained manner.}
\label{fig2}
\end{figure*}
%%%%%%%%%%%%%%%%%%%%%%%%%%%%%%%%%%%%

During a period of time $t_{\rm inj} = 20$~s, a conical jet with an
half-opening angle of 0.2 rad is injected from a spherical, inner boundary
located at $r=5\times 10^8$~cm. The jet luminosity is $L_{\rm jet} = 2 \times
10^{50}$~erg~ s$^{-1}$  (corresponding to a total kinetic energy of
$4 \times 10^{51}$~erg).
Simulations are run with three different values of the jet Lorentz factors:
10, 20, and 30. The jet internal pressure is assumed to be a small fraction ($10^{-5}$) of the rest-mass energy density, giving a  negligible
amount of thermal energy in the jet\footnote{Other authors (e.g. Mizuta \& Ioka 2013) studied the propagation of ``hot'' jets, in which a large fraction of the jet energy is initialized as thermal energy. As most of the jet kinetic energy is dissipated into thermal energy in the post-shock region when the jet moves through the star, the different choices of the initial conditions should produce similar results.}.
 The jet is switched off after
20 seconds, and subsequently the spherical, inner region (from where the jet was injected until that time) becomes part of the computational region.

A computational box with physical size $(L_r,L_z)=(3 \times 10^{14},
3 \times 10^{14})$~cm (along the $r$- and $z$-axis respectively) is resolved
by using a grid with 40 $\times$ 40 cells and 20 levels of refinement,
corresponding to a maximum resolution of $\sim 1.4 \times 10^7$ cm. The jet at the  inner boundary is resolved by $\sim$ 7 cells in the transverse direction. The propagation of the jet is followed for 10$^4$ seconds (time measured
by a stationary observer at the centre of explosion).

As the jet expands, it would be impossible to keep the same resolution during the entire duration of the simulation (as it would require employing $\approx 10^{15}$ cells!). For this reason, we adapt dynamically the grid, refining (i.e., creating four ``sibling'' cells from a parent cell) and derefining (i.e., replacing four sibling cells with one parent, larger cell) the cells as a function of the evolutionary stage of the simulation and the distance $R$ from the origin of the coordinate system. This is done by decreasing the maximum number of levels as $N_{\rm max} = 20 - 3.32 \times \log (t/20)$ for $t> 20$ s, in such a way that the cocoon (whose size increases with time as $R\sim ct$) is resolved by a maximum of approximately $(N_r, N_z) = (2000\times2000)$ cells during its full evolution.

We also run simulations with 22 levels of refinement (within a smaller computational box). Although the details of the jet propagation (i.e. the generation of instabilities at the jet/cocoon interface) depend on resolution, the calculation of the flux and light curve shown in Section \ref{sec:cocoon} do not change by more than 10\% when increasing the resolution from 20 to 22 levels of refinement.

Resolving the density stratification of the wind medium outside the star
requires setting initially a large number of cells which remain unused until the GRB/cocoon shocks arrive at that radius. To make the simulation faster we derefine at the lowest level of refinement all the cells outside a spherical region which is expanding at the speed of light. As this ``virtual'' sphere expands in the environment, the mesh is refined and the values of the density and the pressure are set using the initial conditions of the simulation. In this way, we get an improvement by a factor of $\sim$ 2 in computational time.

%%%%%%%%%%%%%%%%%%%%%%%%%%%%%%%%%%%%%%%%%%%%%%%%%%%%%%%%%%%%%%%%%%%%%%%%%%%%%%
% DYNAMICS
%%%%%%%%%%%%%%%%%%%%%%%%%%%%%%%%%%%%%%%%%%%%%%%%%%%%%%%%%%%%%%%%%%%%%%%%%%%%%%

\section{Dynamics}
\label{sec:dynamics}

%%%%%%%%%%%%%%%%%%%%%%%%%%%%%%%%%%%%
\begin{figure}
\centering
\includegraphics[width=7cm]{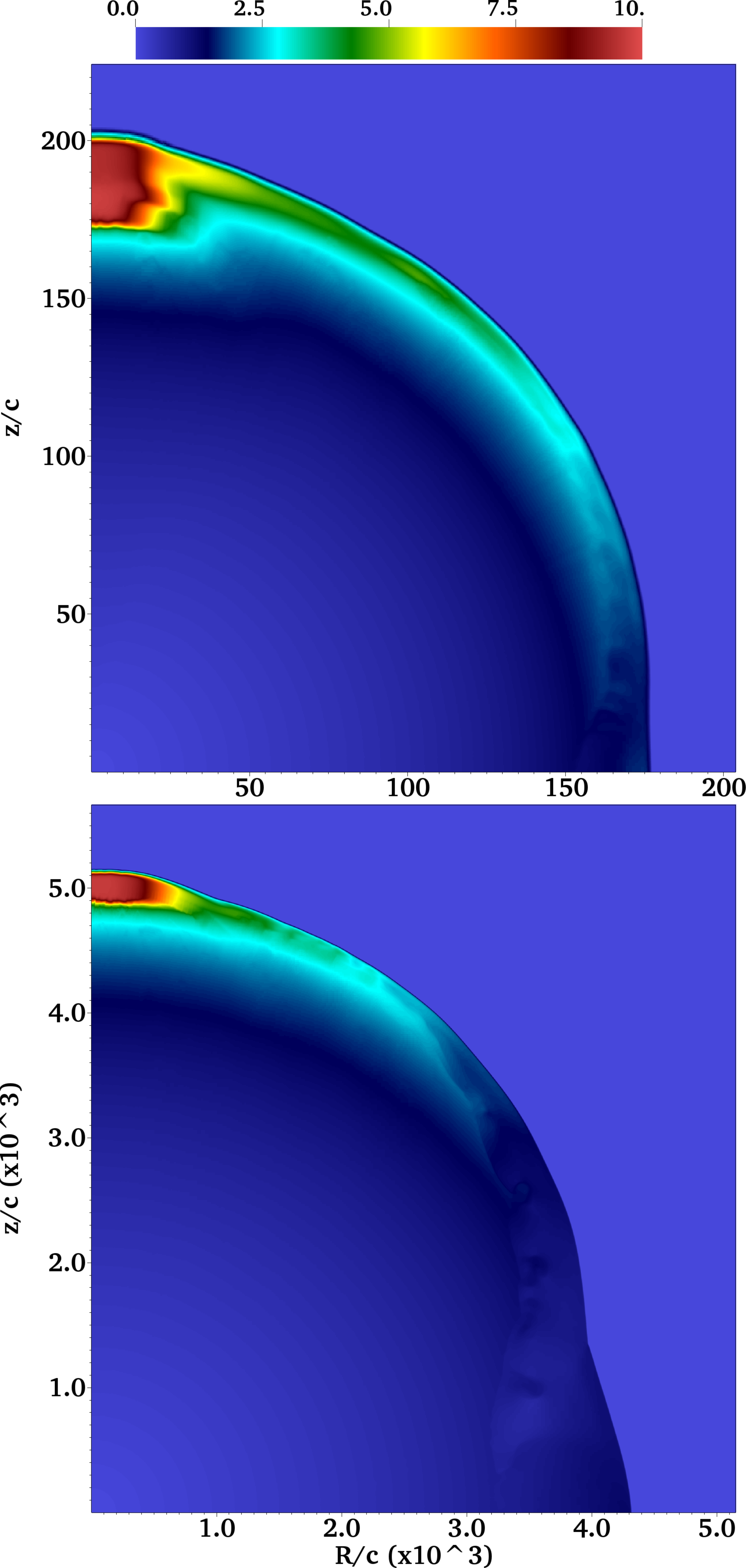}
\caption{
Velocity ($u=\Gamma v/c$) at $t=200$~s (top panel) and  at $t=5000$~s
from the simulation with $\Gamma_{\rm j}=10$ and the E25 progenitor.
}
\label{fig3}
\end{figure}
%%%%%%%%%%%%%%%%%%%%%%%%%%%%%%%%%%%%

The different stages of the dynamical evolution of the GRB/cocoon system\footnote{We describe in this Section only the dynamical evolution of the GRB jet with $\Gamma_{\rm jet}=10$ propagating through the E25 pre-supernova stellar model. The other models present a similar dynamical evolution.} are shown in Figure \ref{fig2}. The jet is injected from a circular boundary (top, left panel of Figure \ref{fig2}). Due to the large stellar density the jet propagates at sub-relativistic speed inside the star, needing about five seconds to get to $r/c \sim$ 0.45 s (top, central panel of Figure \ref{fig2}), which corresponds to an average velocity of $v/c \sim 0.1$. A double shock structure (the ``working surface'') is created at the head of the jet, with a forward shock which accelerates the stellar material, and a reverse shock which decelerates the fast-moving jet material. The hot, dense shocked plasma expands laterally (due to pressure gradients) forming a dense, hot cocoon which helps collimate the jet. In the cocoon, the interface between the shocked jet and the shocked stellar material (the ``contact discontinuity'') is corrugated by the presence of Kelvin-Helmholtz instabilities.

The top, right panel of Figure \ref{fig2} shows the jet breaking out from the star. As the jet arrives at the stellar envelope, the lower entropy stellar material facilitates the lateral expansion of the cocoon, which quickly encloses the star and propagates into the wind medium. The amount of material in the hot
cocoon is then given by the stellar material which has crossed the forward shock when the jet was still inside the star but did not have enough time to move laterally and dissipate its thermal energy inside the star.

The bottom panels of Figure \ref{fig2} show the late evolution (left to right) of the GRB jet/cocoon system.
The cocoon (bottom left panel) is strongly stratified, with a much larger density close to the jet axis and decreasing by 6-7 orders of magnitude at large polar angles. The deceleration phase is clearly visible in the right, bottom panel where Rayleigh-Taylor instabilities form at the interface between the GRB cocoon and the shocked ambient medium along the equatorial plane.
Along the direction of propagation of the GRB jet, the dynamical evolution is more complex. At $t=20$ s the GRB jet is switched-off, and a rarefaction wave moves at the speed of light towards the shock front (visible at $z/c=10$~s and $r/c=0$~s in the left bottom panel). As the rarefaction shock arrives to the head of the jet, the double shock structure gradually forms a thin shell which will move with constant speed up to distances $10-100$ times larger than those simulated here before decelerating.

Figure \ref{fig3} shows the velocity distribution ($u = \Gamma v/c=\Gamma\beta$) at two evolutionary times. Most of the cocoon expands at sub-relativistic speeds, except the region close to the shock front, which moves with larger velocities. The Lorentz factor strongly depends on the polar angle. At $t=200$~s (top panel) the Lorentz factor drops from $\Gamma \sim 5$ (in the region $10^\circ \lesssim \theta  \lesssim 45^\circ$) to sub-relativistic speeds $\Gamma\beta \lesssim 1$ at $\theta\gtrsim 75^\circ$. At $t=5000$~s (bottom panel of Figure \ref{fig3}) the cocoon has slightly decelerated (except for the region $\theta\lesssim 30^\circ$).

\begin{figure}
\centering
\includegraphics[width=8cm]{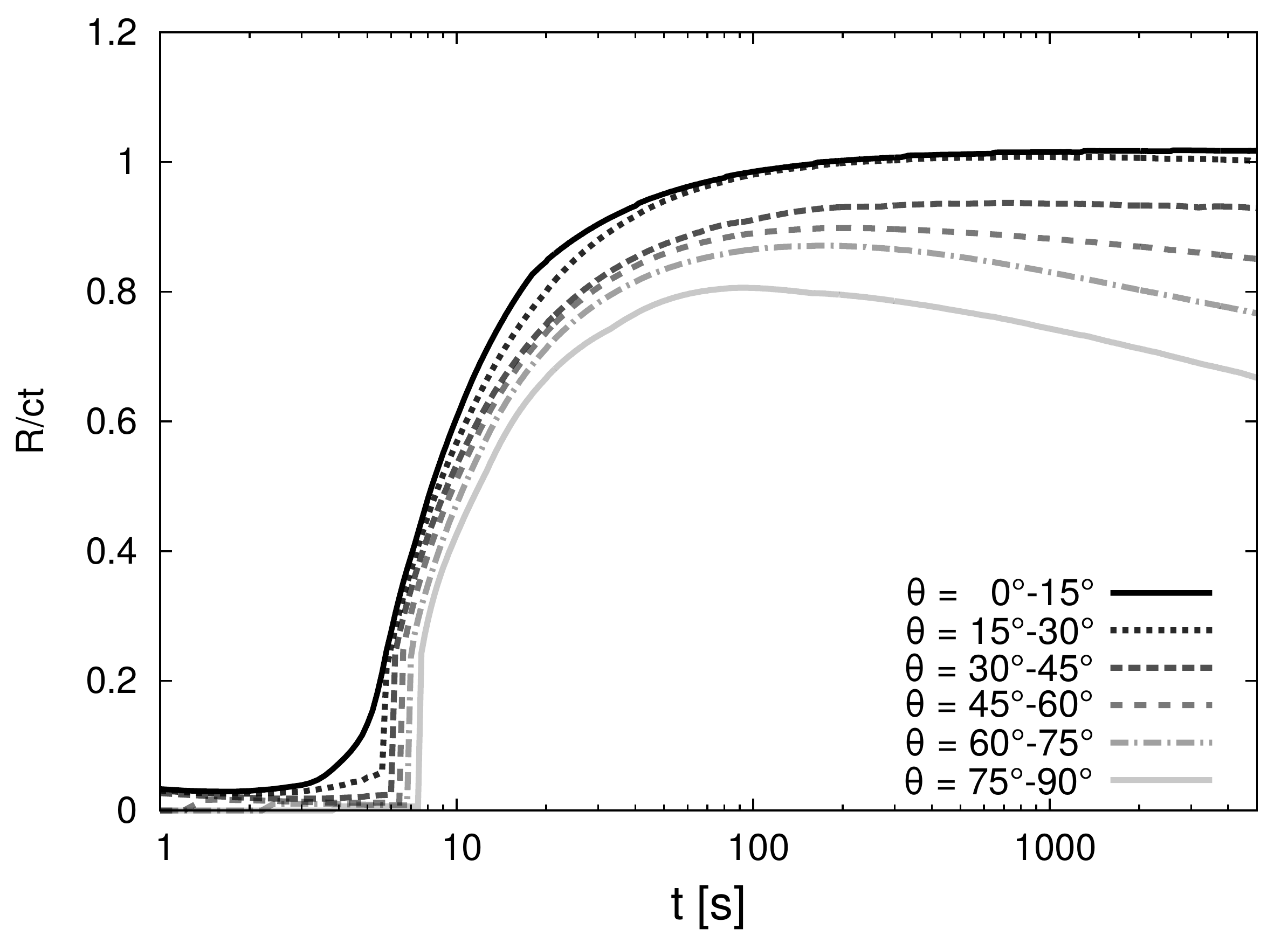}
\caption{Time evolution of the shock average velocity ($R/ct=\int_0^t \beta(t^\prime)  dt^\prime/t$,in a certain range of polar angles (measured from the jet axis). The jet/cocoon system accelerates to relativistic speed in $t\sim 10$ s. For small values of $\theta$ the jet moves at relativistic speeds during the duration of the simulation, while at $\theta\gtrsim 45^\circ$ the cocoon begins to decelerate at $t\gtrsim 100$ s.}
\label{fig4}
\end{figure}

Figure \ref{fig4} shows the time evolution of the shock average velocity
$R/ct$, in the lab frame, as a function of the shock polar angle. As discussed above, the jet moves at non-relativistic speeds while it crosses the star. As it breaks out of the star, it accelerates during $\sim$ 10 seconds
and achieves highly relativistic speeds along the jet axis (with a Lorentz factor equal to the Lorentz factor of the material injected from the inner boundary), and mildly relativistic speeds (with Lorentz factors of $\approx$~2-5, see Figure \ref{fig4}) at large polar angles.
The deceleration phase starts first at larger polar angles.  For instance, the cocoon shock velocity at polar angles $\theta = 75^\circ$-$90^\circ$ decreases from $v \sim 0.8\; c$ at $t=100$~s to $v \sim 0.7\; c$ after $t=10^4$~s.

\begin{figure}
\centering
\includegraphics[width=8cm]{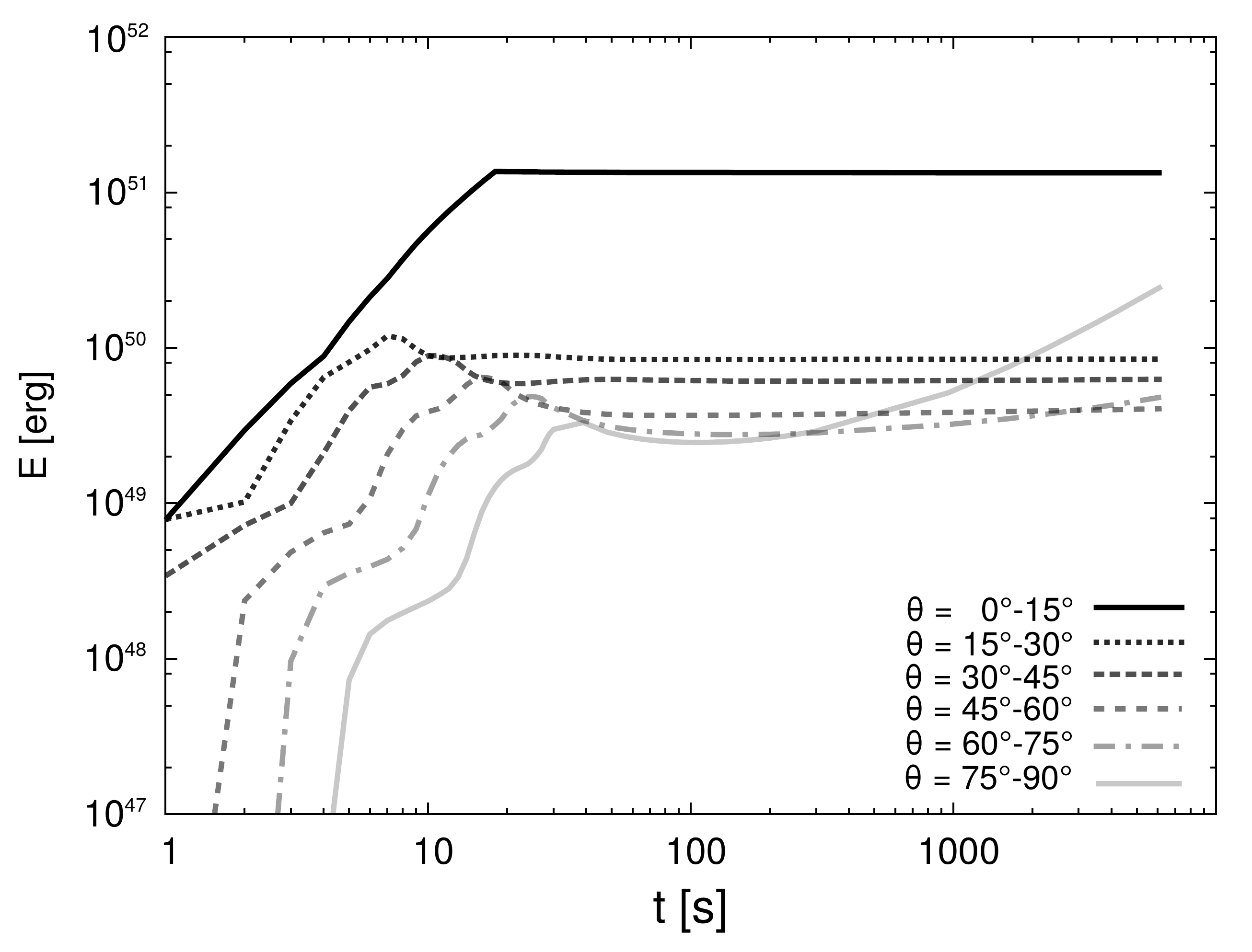}
\caption{Jet/cocoon energy in different intervals of $\theta$ as a function of time. Most of the energy remains collimated in the jet half-opening angle $\theta_{\rm jet}\approx 10^\circ$ and increases linearly with time over the
duration of injection of relativistic material from the central engine (assumed here to be 20 seconds).
}
\label{fig5}
\end{figure}

Figure \ref{fig5} shows the energy distribution as a function of polar angles. The energy increases during the first 20 seconds (consistently with the duration of injection of the jet from the inner boundary). Most of the energy remains collimated in a conical region of angular size $\theta\lesssim 15^\circ$ during the full duration of the numerical simulation. A smaller amount of energy ($10^{49}$-$10^{50}$ erg) is present in the portion of the cocoon moving at larger angles. While along the jet axis most of the energy is concentrated in a small region around the shock, at larger polar angles a significant fraction of the energy is distributed through all the cocoon's volume. Therefore, the
deceleration radius observed in the numerical simulations is smaller than
one inferred by using the expression $R/c\sim E v_w/(\dot{M}_w \Gamma^2 c^3) \approx 6\times10^3 E_{50}/\Gamma_1^2$ s.

It is very challenging, even with the excellent computational resources available these days, to run numerical simulations up to several times $10^{14}$~cm in 3D. 
The effect of 3D versus 2D simulations has been recently studied by, e.g., \citet{gottlieb18, harrison18}. These authors showed that the presence of asymmetries in the jet (i.e., along the direction perpendicular to the jet axis) reduces the velocity of the cocoon thereby making the cocoon dimmer, with larger differences depending on the amount of asymmetry present in the jet\footnote{A 3D simulation with cylindrical symmetry would obviously recover the results of the 2D axisymmetric simulations unless the code is not, by construction, preserving symmetries} (i.e. if the jet is precessing, wiggling and so on).

%%%%%%%%%%%%%%%%%%%%%%%%%%%%%%%%%%%%
%%%%%%%%%%%%%%%%%%%%%%%%%%%%%%%%%%%%

%%%%%%%%%%%%%%%%%%%%%%%%%%%%%%%%%%%%
%%%%%%%%%%%%%%%%%%%%%%%%%%%%%%%%%%%%

%%%%%%%%%%%%%%%%%%%%%%%%%%%%%%%%%%%%%%%%%%%%%%%%%%%%%%%%%%%%%%%%%%%%%%%%%%%%%%
% RADIATION
%%%%%%%%%%%%%%%%%%%%%%%%%%%%%%%%%%%%%%%%%%%%%%%%%%%%%%%%%%%%%%%%%%%%%%%%%%%%%%

\section{Cocoon luminosity and spectrum}
\label{sec:cocoon}

\subsection{Radiative diffusion}

The average optical thickness of each cell in our simulation is
\begin{eqnarray}
 \delta\tau &\sim& \sigma_T M_{\rm c}/(2\pi m_p N_{\rm cells} R^2) \\
 &\sim& 7 (M_{\rm c}/10^{-4} M_{\odot})
R_{12}^{-2}
\end{eqnarray}
where $N_{\rm cells}=2000$ is the maximum number of cells in the simulations
along the radial and transverse directions
(the cocoon terminal Lorentz factor for mass $10^{-4}M_{\odot}$
and energy of 10$^{51}$ erg is $\sim5$).
However, the whole cocoon remains optically thick
along the polar axis for $\sim 10^3$ s when the cocoon reaches
$R\sim 3\times10^{13}$ cm. The cocoon is nearly isobaric throughout the
volume (except near the surface) and the pressure
is dominated by radiation. Therefore, the fractional change to the thermal
energy of an interior cell in a dynamical time is of order
$(N_{\rm cells}\,\delta\tau)^{-1} \ll 1$. Therefore, for cells below the photosphere, diffusive radiation has a negligible effect
on the dynamics, and post-processing of the hydro-simulation output to
calculate cocoon radiation is expected to yield a reasonably accurate result at least for small polar angles $\lesssim 45^{\rm o}$. At large polar angles, the cocoon structure is affected by the loss of thermal energy at late evolutionary times. A full special relativistic radiation hydrodynamics calculation is necessary to properly compute the cocoon radiation seen by an observer located off-axis. In this paper, we only consider the case where the observer is on-axis at $\theta_{\rm obs}=0$.

Although radiative diffusion does not change the hydrodynamical evolution of the system,
it strongly modifies the temperature structure of the cocoon surface and the resulting cocoon radiation. In our post-process procedure, radiative diffusion is coupled to the hydro-simulations as follows.
Given a snapshot of the numerical simulation, the flux diffusing across the boundaries of each cell is calculated by employing the {\it flux-limited diffusion} approximation in the comoving frame of each cell
\begin{equation}
 \vec F^{\prime} = - \frac{c \lambda}{k_{\rm es} \rho^{\prime}} \nabla^{\prime} e^{\prime}_{\rm rad}\;,
 \label{flux}
\end{equation}
where $e^{\prime}_{\rm rad}$ is the (comoving) radiation energy density (defined as $e^{\prime}_{\rm rad} = a T^{\prime,4}$, with the proper temperature defined in equation \ref{eq:et}) of nearby cells, $\rho^{\prime}$ is the matter  density, %and $k_{\rm es} = 0.2\; (1+X)$  cm$^2$  g$^{-1}$
and $k_{\rm es} = 0.2$  cm$^2$  g$^{-1}$
is the electron scattering opacity for fully ionised helium (or heavier elements).
The parameter $\lambda$ is the \emph{flux limiter}, given by \citep{levermore81}:
\begin{equation}
 \lambda = \frac{2+\xi}{6+3\xi+\xi^2} \;,
\end{equation}
where $\xi$ is a dimensionless quantity defined as
\begin{equation}
 \xi = \frac{|\nabla^{\prime} e^{\prime}_{\rm rad}|}{k_{\rm es}\ \rho^{\prime} e^{\prime}_{\rm rad}} \;.
\end{equation}
Equation (\ref{flux}) gives the correct scaling in the optically thin and optical thick limits, although it represents only an approximation of the intermediate case. In an optically thick medium $\xi \ll 1$ and $\lambda= 1/3$, equation (\ref{flux}) reduces to
\begin{equation}
 \vec F^{\prime} = - \frac{c}{3 k_{\rm es} \rho^{\prime}} \nabla^{\prime} e^{\prime}_{\rm rad}\;,
\end{equation}
while in the optically thin limit $\xi \gg 1$ and the flux is limited to %
\begin{equation}
 \vec F^{\prime} = - c \vec{e}_{\rm rad}^{\prime} \;,
\end{equation}
where $\vec{e}_{\rm rad}^{\prime} = \nabla e_{\rm rad}^{\prime}/|e_{\rm rad}^{\prime}|$.

Once the flux is computed using equation (\ref{flux}), the effect of the radiation
is included in the energy equation by solving the following equation
\begin{equation}
 \frac{\partial e^{\prime}_{\rm gas+rad}}{\partial t^{\prime}} = - \nabla^{\prime} \cdot F^{\prime}_{\rm rad} \;,
\end{equation}
which in cylindrical coordinates takes the following form
\begin{eqnarray}
 \frac{\partial e^{\prime}_{\rm gas+rad}}{\partial t^{\prime}}
 =
 - \frac{1}{r^{\prime}} \frac{\partial (r^{\prime} F^{\prime}_r)}{\partial r^{\prime}}
 -  \frac{\partial^{\prime} F^{\prime}_z}{\partial r^{\prime}} =
 \nonumber \\
 =
- \frac{1}{r^{\prime}} \frac{\partial}{\partial r^{\prime}}
\left(  r^{\prime}
  \frac{c \lambda}{\rho^{\prime} k_{\rm es}} \frac{\partial e^{\prime}_{\rm rad}}{\partial r^{\prime}}
\right)
-
 \frac{\partial }{\partial z^{\prime}}
\left(
  \frac{c \lambda}{\rho^{\prime} k_{\rm es}} \frac{\partial e^{\prime}_{\rm rad}}{\partial z^{\prime}}
\right)\;,
 \end{eqnarray}
and is discretized as
\begin{eqnarray}
e_{{\rm gas+rad},i,j}^{\prime,n+1}
& = &
e_{{\rm gas+rad},i,j}^{\prime,n}\nonumber \\
&-&
\frac{\Delta t^{\prime,n}}{r^{\prime}_i \Delta r^{\prime}_{i,j}}
\left(
r^{\prime}_{i^+}  \nu_{i^+,j} \frac{\Delta e^{\prime}_{i^+}}{\Delta r^{\prime}_{i,j}}
- r^{\prime}_{i^-} \nu_{i^+,j} \frac{\Delta e^{\prime}_{i^-}}{\Delta r^{\prime}_{i,j}}
\right) \nonumber \\
&-&
 \frac{\Delta t^{\prime,n}}{\Delta z^{\prime}_{i,j}}
\left(
  \nu_{i,j^+} \frac{\Delta e^{\prime}_{j^+}}{\Delta z^{\prime}_{i,j}}
- \nu_{i,j^-} \frac{\Delta e^{\prime}_{j^-}}{\Delta z^{\prime}_{i,j}}
\right)\;,
\label{eq:energy}
\end{eqnarray}
where
$i^{\pm}=i \pm 1/2$, $\nu_d = \frac{c \lambda}{\rho k_{\rm es}}$, $\Delta e^{\prime}_{i^+} = e^{\prime}_{{\rm rad}, i+1,j}-e^{\prime}_{{\rm rad},i,j}$ and the indexes $i,j$ refer to the cell $(i,j)$ of the computational grid, with $i=1, \dots, N_r$, $j=1, \dots, N_z$ ($N_r$ and $N_z$ being the total number of cells along the $r$ and $z$ directions respectively). Finally, we notice that the stability condition restricts the time step to $ \Delta t \leq  (\Delta x)^2/4 \nu_d$.

\subsection{Calculation of luminosity and spectrum}

For each simulation, we save a large number of snapshots (typically 10$^3$).
After post-processing each snapshot with the radiative diffusion algorithm
described above and assuming that the observer is located along the $z$-axis,
we determine the position of the photosphere for each snapshot, defined as
the surface corresponding to an optical depth $\tau=1$, i.e.
\begin{equation}
  \label{eq:1}
  \int_z^\infty \kappa_{\rm es} \rho^\prime \Gamma \d z = 1.
\end{equation}
For each cell $(j,k)$ located on the photosphere we compute the specific intensity at the location of the observer and assign it to the corresponding observing time, related to the simulation time by
\begin{equation}
   t_{\rm obs} = t_k - z_j/c,
\end{equation}
where $z_j$ is the distance from the equatorial plane and $t_k$ is the
simulation time in the lab frame. The duration of the snapshot $k$ can
be defined as $\Delta t_k = (t_{k+1}-t_{k-1})/2$, and the flux in the
observer's frame lasts for $\Delta t_{\rm obs} = (1-\beta_j \cos \alpha)
\Delta t_k$, where $\alpha$ is the angle between the velocity vector and
the $z$-axis and $\beta_j$ is the photosphere velocity.
As our purpose is to compute the specific and total luminosity, we neglect
cosmological effects and take $z=0$.

To compute the flux we add up the contributions from all cells at the photosphere \citep[see, e.g.,][]{decolle12a}
\begin{equation}
   dF_{\nu_{\rm obs}}^{\rm obs} = I_\nu^{\rm obs} \cos \theta_{d} d\Omega_{d} \approx I_\nu^{\rm obs} \frac{dS_\perp}{d^2}\;,
\end{equation}
where d is the distance to the source, $\theta_d$ is the angle between the
line joining the observer to the centre of the star and the line from
the observer to the volume element from which radiation is being
considered, and $d\Omega_{d} = dS_\perp/d^2$ is the differential solid
angle of the cell as viewed by the observer. As the angular size of the
source according to the observer is negligibly small, it is justified to
take $\cos \theta_{d}\approx 1$.
The specific intensity in the comoving frame is obtained by assuming a blackbody spectrum
\begin{equation}
\label{eq:intensity}
  I^{\prime}_{\nu^{\prime}} \approx \frac{F^{\prime}}{\pi} \frac{B_{\nu^{\prime}}(T^{\prime})}{\int B_{\nu^{\prime}}(T^{\prime}) d\nu^{\prime}} = \epsilon B_{\nu^{\prime}}(T^{\prime})\;,
\end{equation}
where $F^\prime$ is given by equation (\ref{flux}).

The comoving-frame temperature $T^{\prime}$ is computed from the values of energy (obtained at each timestep by using equation \ref{eq:energy}) and density by inverting the equation
\begin{equation}
   e^{\prime} = aT^{\prime 4} + \frac{3}{2} \frac{k_B\rho^{\prime} T^{\prime}}{\mu m_p},
  \label{eq:et}
\end{equation}
where $\mu = 4/3$ is the mean molecular weight. The specific intensity in the observer's frame is given by
\begin{equation}
  I^{\rm obs}_{\nu_{\rm obs}} = I^{\prime}_{\nu^{\prime}} \left(\frac{\nu_{\rm obs}}{\nu^{\prime}}\right)^3, \quad \nu_{\rm obs} = {\nu^{\prime} \over \Gamma (1 - \beta \cos\alpha)}.
\end{equation}

We employ the fact that the blackbody photon distribution function is invariant under Lorentz transformation, with the temperatures in the two frames related by the following relation
\begin{equation}
   T_{\rm obs} = \frac{T^{\prime}}{\Gamma (1-\beta\cos \alpha)}.
\end{equation}

We compute the total flux in the observer frame (see, e.g.,
\citealt{decolle12a}) by integrating over the entire photosphere
\begin{equation}
  F_{\nu_{\rm obs}}^{\rm obs} = \frac{1}{d^2}  \int d S_\perp \epsilon B_{\nu_{\rm obs}}(T_{\rm obs}) \delta \left( t_{\rm obs} -  t + \frac{z}{c}
 \right)\;.
\end{equation}
Integrated and averaged over the time interval $\Delta t_{\rm obs}$, this equation reduces to
\begin{equation}
  F_{\nu_{\rm obs}}^{\rm obs}(t_{\rm obs}) = \frac{1}{d^2 \Delta t_{\rm obs}} \sum_{j,k} \epsilon B_{\nu_{\rm obs}}(T_{\rm obs,j,k}) \d S_{\perp,j,k} \d t_{j}\;,
\end{equation}
where the sum extends over all $j$-th snapshots (one at each lab frame time $t_j$) and $k$-th cells respectively. The luminosity in the observer's  frame is
\begin{equation}
    L_{\nu_{\rm obs}}(t_{\rm obs}) = 4 \pi d^2 F_{\nu_{\rm obs}}^{\rm obs} \;.
\end{equation}
%

%%%%%%%%%%%%%%%%%%%%%%%%%%%%%%%%%%%%
\begin{figure}
\centering
\includegraphics[width=8cm]{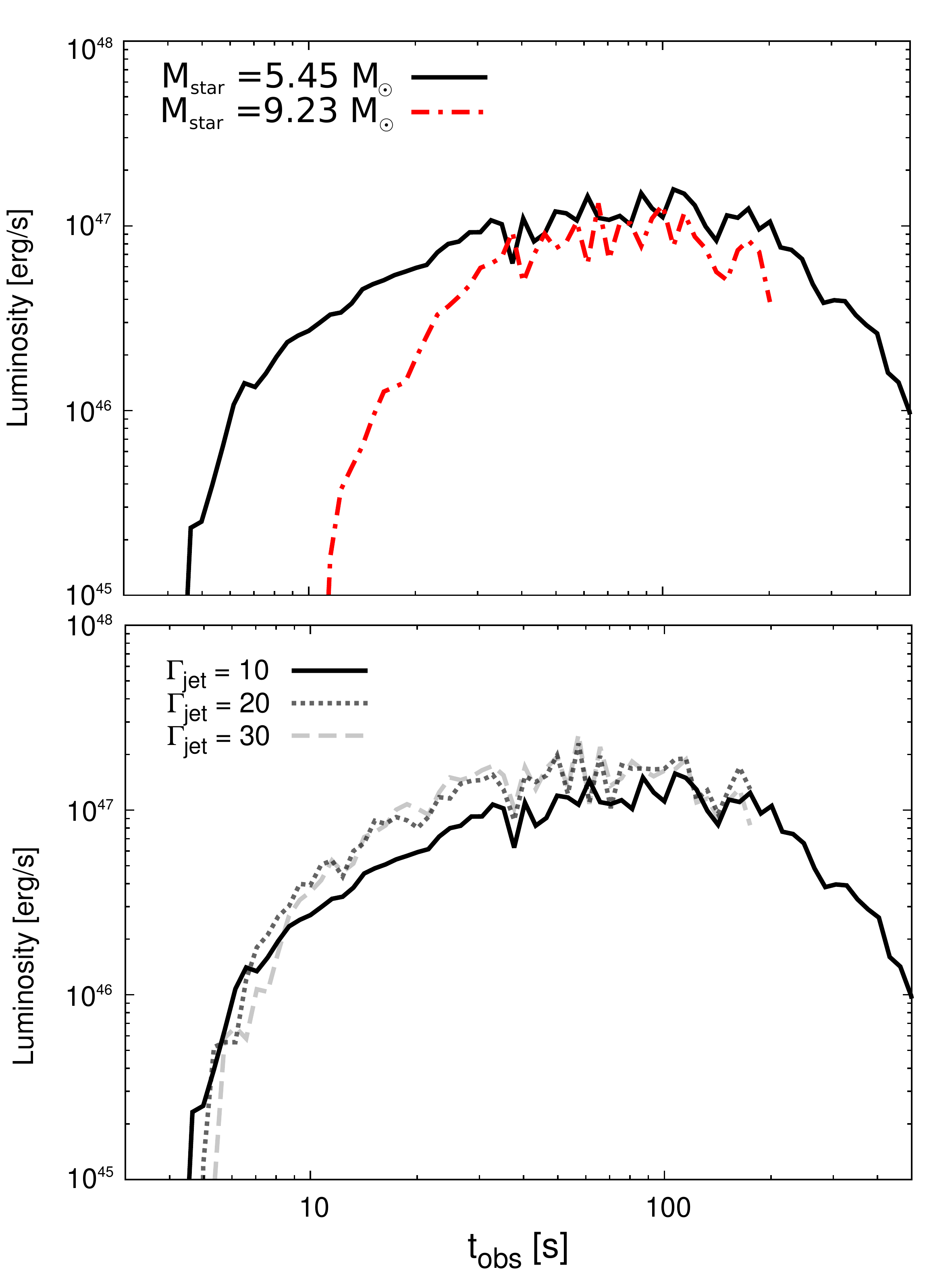}
\caption{Bolometric luminosity lightcurve (in the observer frame) computed by post-processing the results of the numerical simulations. \emph{Top}: This panel shows the effect of different stellar structures on the lightcurve (see figure \ref{fig1}).
\emph{Bottom}: This panel shows that the lightcurve is nearly independent on the Lorentz factor of the GRB jet.
Depending on the duration of the simulation and the jet Lorentz factor, some of the simulations have luminosities spanning a shorter time.}
\label{fig6}
\end{figure}
%%%%%%%%%%%%%%%%%%%%%%%%%%%%%%%%%%%%

%%%%%%%%%%%%%%%%%%%%%%%%%%%%%%%%%%%%
\begin{figure}
\centering
\includegraphics[width=8cm]{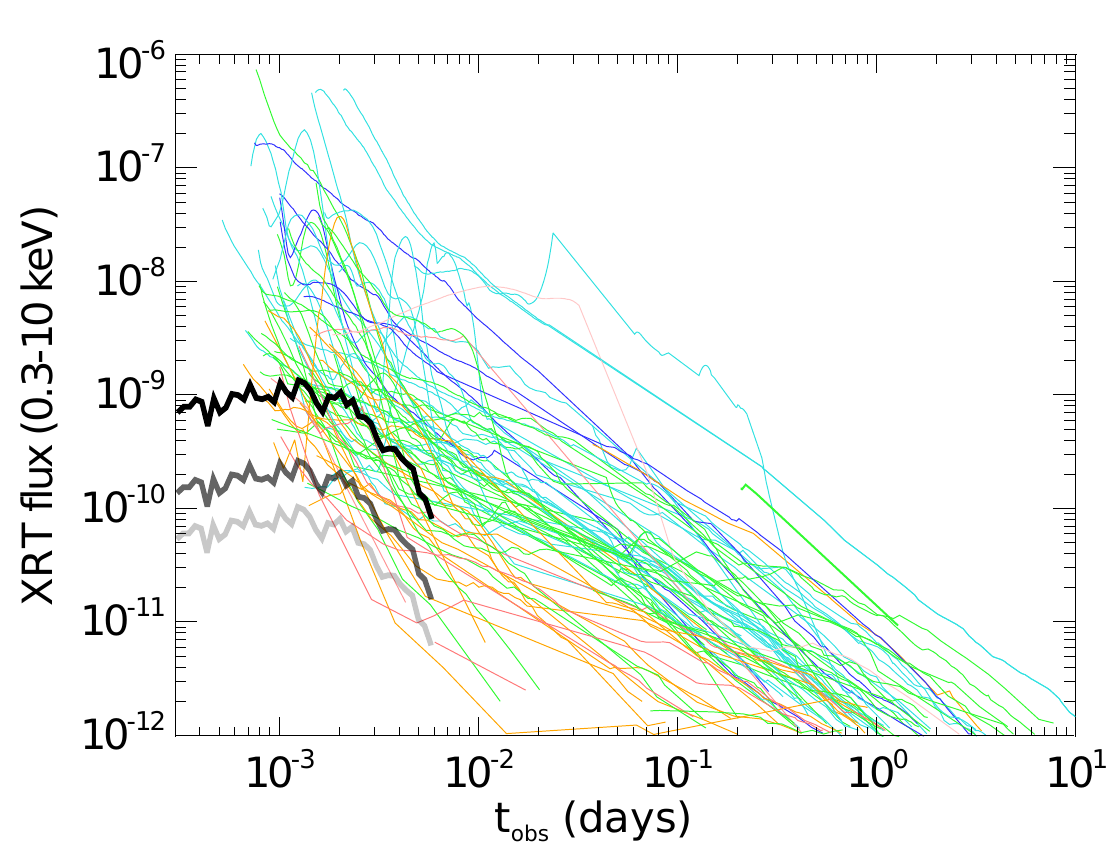}
\caption{X-ray flux computed from the numerical simulations ($\Gamma_{\rm jet}=10$ and M$_{\rm star} = 5.45$ M$_\odot$) at redshifts  $z=0.2,0.4,0.6$ (black to grey curves), compared with a sample of GRBs with $z < 0.8$ \citep[adapted from][]{perley14}. The afterglow curve colours correspond to isotropic energies
$E_{\gamma,{\rm iso}} < 10^{51}$~ergs (red),
$10^{51}-10^{52}$~ergs (orange),
$10^{52}-10^{53}$~ergs (green),
$10^{53}-10^{54}$~ergs (cyan),
$>10^{54}$~ergs (blue). The figure shows that the cocoon X-ray flux is of the same order as the X-ray flux of a typical GRB during the plateau phase.
}
\label{fig7}
\end{figure}
%%%%%%%%%%%%%%%%%%%%%%%%%%%%%%%%%%%%

\subsection{Results}

We present simulation results for cocoon luminosity and spectra for two
different GRB progenitor star models (E25 and 12TH of \citealt{heger00, woosley06b}).
For the model E25, three different jet Lorentz factors (10, 20 and 30) are considered.

Figure \ref{fig6} shows the bolometric cocoon luminosity computed with the method described in the previous section.
The luminosity increases quickly to $\sim 10^{45}-10^{46}$ erg/s as the jet breaks out of the star and increases to values of $\sim 10^{47}$ erg/s after 100 s (in the observer's frame) before dropping as $\sim t_{\rm obs}^{-3}$ for larger observing time.

The X-ray lightcurve of a sample of GRBs \citep{perley14} is compared in figure \ref{fig7} with the X-ray flux emitted by the cocoon.
The cocoon energy is a fraction of the energy deposited by the jet into the star (i.e. the jet luminosity integrated over the time needed for the jet to cross the star), i.e. $\sim$ 1/100-1/10 of the GRB total energy (10$^{49}$-10$^{50}$~erg, see Figure \ref{fig5}). The GRB kinetic luminosity (imposed as 10$^{50}$ erg/s in our simulations) is about three orders of magnitude larger than the cocoon luminosity ($\sim$ 10$^{47}$ erg/s). The cocoon luminosity lasts for a much longer period of time ($\sim$ several hundred of seconds, see \ref{fig6}), and the total energy emitted by the cocoon is about $\sim 5 \times 10^{49}$ erg. 
The cocoon X-ray luminosity is much smaller than GRB jet afterglow luminosity for the initial 30-50 s. Subsequently, as the GRB jet X-ray lightcurve undergoes a very steep decline for a few minutes after the end of the prompt phase,  the cocoon luminosity becomes of the same order or slightly lower than the afterglows of GRBs with low isotropic energies ($E_{\rm iso} < 10^{52}$~erg), and about one order of magnitude lower than GRBs with large isotropic energies ($E_{\rm iso} > 10^{52}$~erg).
This can be seen by comparing the cocoon luminosity  with e.g. the XRT flux for GRBs in the 10$^{51}$-10$^{53}$~erg range (see Figure \ref{fig7}).
Thus, the cocoon radiation should be detectable at least for some GRBs and can be identified by its distinct thermal spectrum.

Figure \ref{fig6} also shows that the stellar structure is the key parameter which determines the cocoon luminosity. When a more extended stellar model is considered, the lightcurve shows a peak luminosity slightly lower and is in general dimmer at all observing times. As the jet break-out happens at later times, the luminosity increases on a longer timescale.
Finally, increasing the jet Lorentz factor (while keeping fixed the total energy of the jet) does not produce a large change in luminosity (see the bottom panel of Figure \ref{fig6}).
This result is consistent with the cocoon energy being determined by the jet ram pressure/luminosity (which is the same in all models considered here), and on the time it takes for the jet to carve out a polar cavity through the GRB progenitor star, and not on the jet Lorentz factor (the speed at which the jet moves through the star is a weak function of $\Gamma_j$).

\begin{figure}
\centering
\includegraphics[width=8cm]{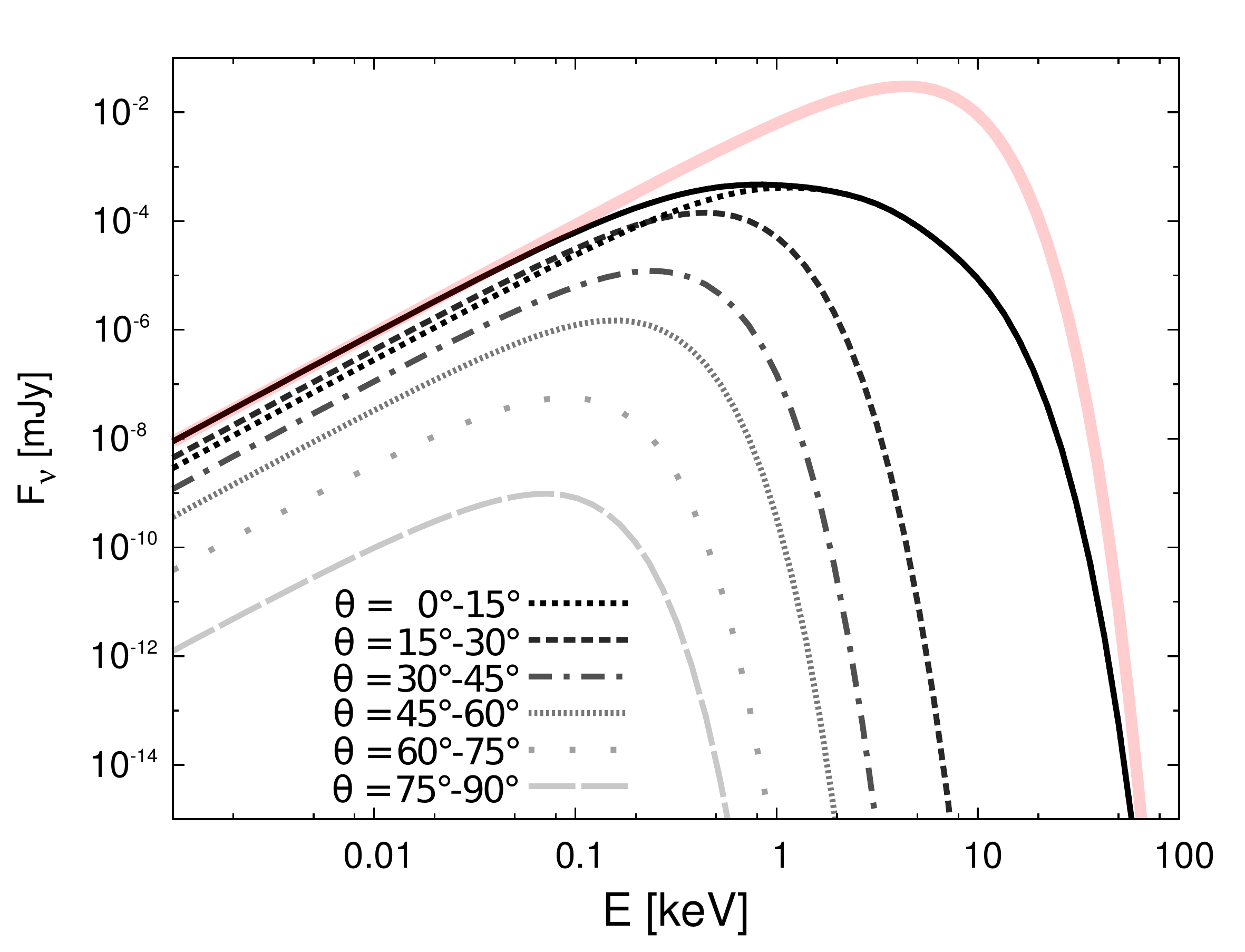}
\caption{Spectra (F$_\nu$) as a function of the polar angle $\theta$ (measured in the lab frame) at $t_{\rm obs}=30$ s for an observer located on-axis. Most of the emission comes from the region close to the jet axis. The black, full line represents the  total flux (computed integrating over the polar direction).
For comparison, a black body spectrum is also shown (pink curve). }
\label{fig8}
\end{figure}

Figure \ref{fig8} shows the angular contribution to the spectrum at
$t_{\rm obs} = 30$ s for the E25, $\Gamma_{\rm jet}=10$ model. The spectrum peaks
in the X-ray band were most of the energy is emitted. Actually, a plot of
the X-ray luminosity (in the 0.3--10 keV band, which corresponds to
Swift XRT energy coverage) would be indistinguishable from the
bolometric luminosity shown in Figure \ref{fig6}.
The spectrum is nearly thermal (a blackbody curve is also shown in the Figure
for comparison), with a low-frequency spectrum $F_\nu\propto \nu^2$ and an exponential decay at large frequencies, and wider than a blackbody spectrum near the peak. Due to relativistic beaming, the flux is dominated by the emission at small angles ($\theta \lesssim 30^\circ$). Angles larger than $45^\circ$ produce a negligible flux when the jet is observed on-axis (they would dominate the cocoon's emission for off-axis GRBs).

\begin{figure}
\centering
\includegraphics[width=8cm]{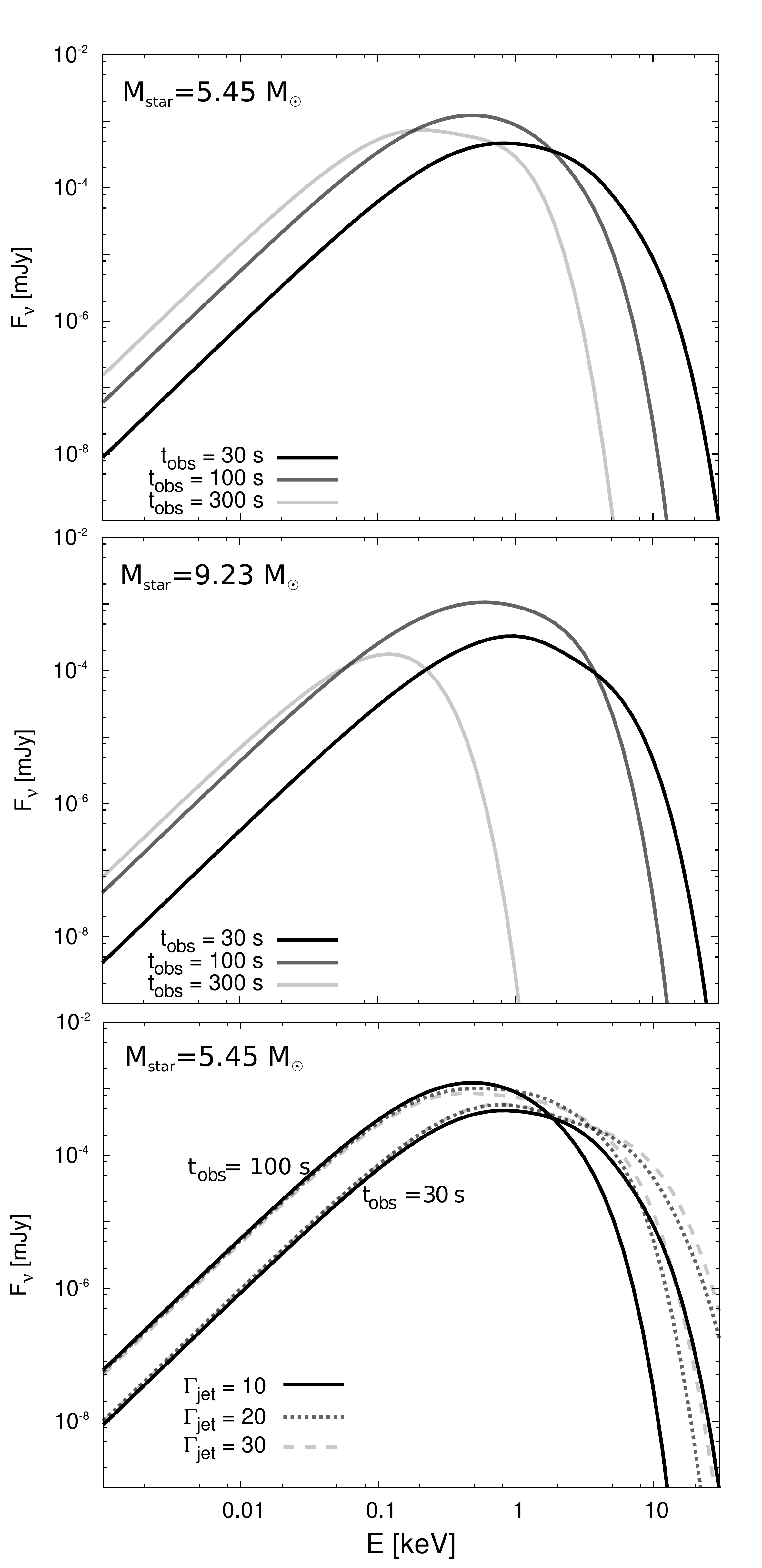}
\caption{\emph{Top, centre panels}: Spectra (F$_\nu$) at $t_{\rm obs}=30, 100, 300$ s for the two pre-supernova stellar models considered (see Figure \ref{fig1}). \emph{Bottom panel}: Spectra at $t_{\rm obs}=30, 100$ s for different jet Lorentz factors.}
\label{fig9}
\end{figure}

\begin{figure}
\centering
\includegraphics[width=8cm]{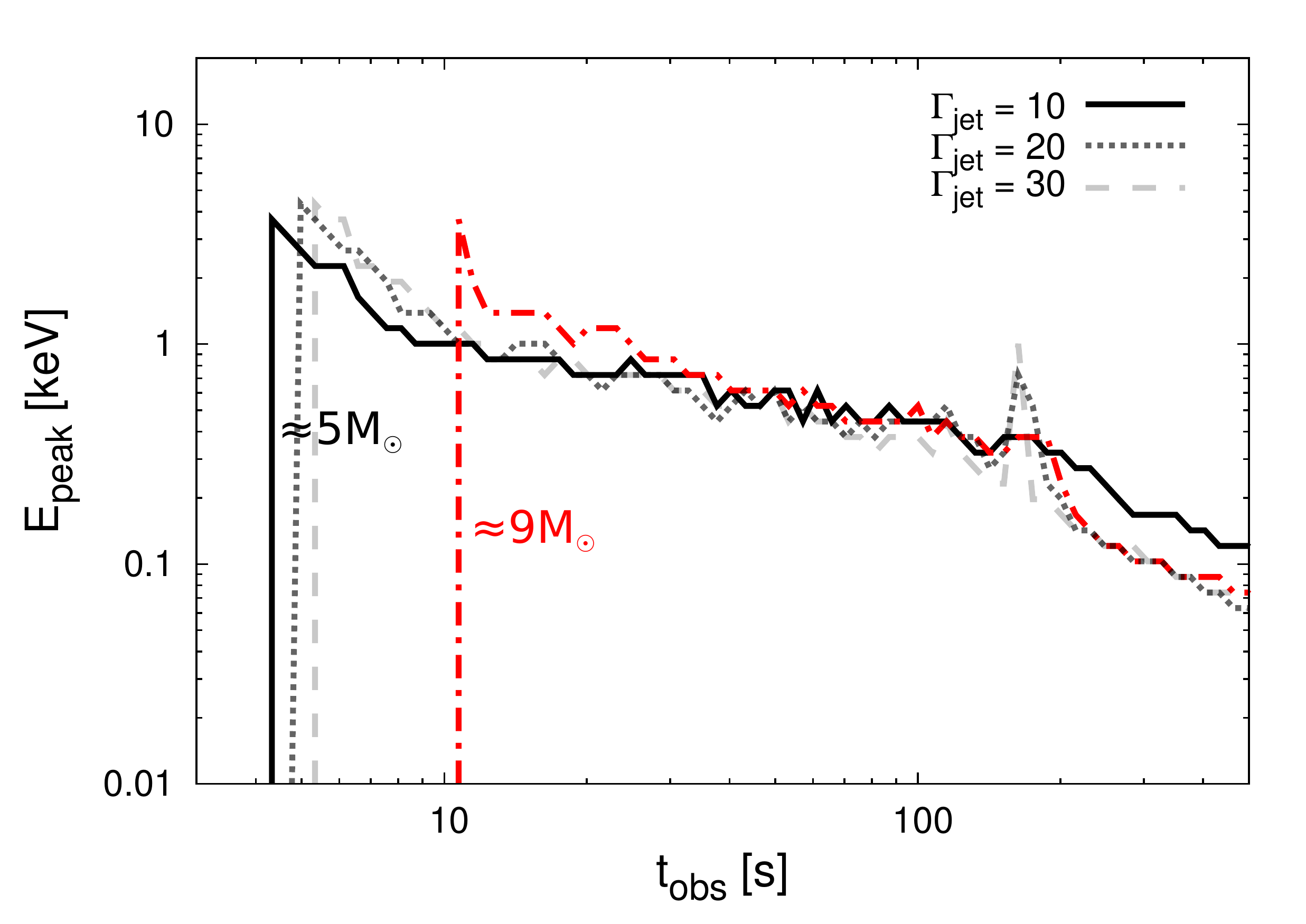}
\caption{Time evolution (in the observing frame) of the energy peak of the quasi-thermal spectrum for the two progenitor stars and jet Lorentz factors considered in the numerical simulations.}
\label{fig10}
\end{figure}

Figures \ref{fig9} and \ref{fig10} show the time evolution of the spectrum and energy peak respectively at $t_{\rm obs} = 30, 100$~s.
The cocoon spectrum for a more extended stellar model (Figure \ref{fig9},
central panel) is less similar to a blackbody, and its evolution with time is
more rapid. Also shown in the Figure \ref{fig9} are spectra for three
different values of jet Lorentz factor; the spectrum below $\lesssim$~1 keV
is nearly independent of $\Gamma_{\rm jet}$, however, at higher energies the flux
is larger for cocoons formed by jets with larger $\Gamma_{\rm jet}$.

%%%%%%%%%%%%%%%%%%%%%%%%%%%%%%%%%%%%
%%%%%%%%%%%%%%%%%%%%%%%%%%%%%%%%%%%%

%%%%%%%%%%%%%%%%%%%%%%%%%%%%%%%%%%%%
%%%%%%%%%%%%%%%%%%%%%%%%%%%%%%%%%%%%

%%%%%%%%%%%%%%%%%%%%%%%%%%%%%%%%%%%%
%%%%%%%%%%%%%%%%%%%%%%%%%%%%%%%%%%%%

%%%%%%%%%%%%%%%%%%%%%%%%%%%%%%%%%%%%
\subsection{Observations of thermal X-ray radiation}

Several groups searched for thermal X-ray emission after the prompt
$\gamma$-ray phase \citep[e.g.][and references therein]{friis13, starling12, sparre12}.
\citet{starling12} analysed spectra for 11 GRBs
and found thermal X-ray signal for three of these (GRBs 060218, 090618
and 100316D) which are known to be associated with supernovae.
In 4 additional cases the spectral fit is claimed to improve when a thermal
component is included; these 4 cases are also possibly
associated with supernovae. \citet{sparre12} analysed Swift/XRT
data for 190 GRBs and identified 6 bursts with possible blackbody
components\footnote{There
might have been many more bursts with a thermal afterglow component which
the analysis of \citet{sparre12} could not identify due to the
uncertainty associated with hydrogen column density and the absorption of
soft X-ray radiation.}
(GRBs 061021, 061110A, 081109, 090814A, 100621A and 110715A). The three new
candidates identified by \citet[][GRBs 061021, 061110A and
090814A]{sparre12} are not known to have an associated supernova.
The thermal components in these 6 cases were found
during the steep decline or the plateau phase of the X-ray lightcurve.

One of the models suggested for the thermal component in GRB
afterglow lightcurves is that it is the radiation associated with
the supernova shock breaking through the surface of the progenitor star.
However, several papers \citep{ghisellini07, li07, chevalier08}
point out a problem with this scenario, and that is that the thermal
luminosity is too large to be consistent with the shock-breakout model
for GRB 060218; the thermal luminosity for the
other five cases are even larger than GRB 060218 \citep{sparre12}.

The observed luminosity for the thermal component of these GRB afterglows
is between 10$^{47}$ and 10$^{50}$ erg s$^{-1}$, and the spectral
peak is at $\sim 0.5$ keV \citep{sparre12}. These values
are consistent with the expectations of cocoon radiation (see Figs.
\ref{fig6} and \ref{fig10}). Measurement of the
evolution of the thermal component's peak temperature and luminosity in
the future and its comparison with the cocoon simulation would make
it possible to draw a firm conclusion regarding the origin of the radiation.
If confirmed to be originated by the cocoon associated with the relativistic jet,
then that could be used to determine the progenitor star radius and density structure.

%%%%%%%%%%%%%%%%%%%%%%%%%%%%%%%%%%%%%%%%%%%%%%%%%%%%%%%%%%%%%%%%%%%%%%%%%%%%%%
% INVERSE COMPTON
%%%%%%%%%%%%%%%%%%%%%%%%%%%%%%%%%%%%%%%%%%%%%%%%%%%%%%%%%%%%%%%%%%%%%%%%%%%%%%

\section{Inverse-Compton scattering of cocoon radiation by the relativistic jet}
\label{sec:ic}

As thermal photons from the cocoon diffuse out from the photosphere and
run into the relativistic jet, they can get inverse-Compton (IC)
scattered by electrons to much higher energies. This will produce a
flash of high energy photons when the faster jet surpasses the cocoon
photosphere \citep{kumar14}.  Because of the relativistic beaming
along the direction of cocoon local
velocity vector, only photons emitted within an angular
patch of size $\sim\Gamma_{\rm c}^{-1}$ ($\Gamma_{\rm c}$ being the
cocoon Lorentz factor) surrounding the jet can interact with electrons
in the jet and contribute to IC luminosity.
Since the angle between the jet axis and photons' moving direction is of  order $\Gamma_{\rm c}^{-1}$, IC scatterings boost cocoon photon energy
to $\sim kT_{\rm c} (\Gamma_{\rm j}/\Gamma_{\rm c})^2 \sim \Gamma_{\rm j, 2.5}^2/\Gamma_{\rm c, 1}^2$ MeV when
electrons are cold ($\gamma_{\rm e} \sim 1$) in the jet comoving frame.

We present in this section the IC lightcurves and spectra computed by
post-processing the results of the hydrodynamical simulations.
To calculate the IC emission in real GRBs, we consider the following two cases. (i) In the {\it Early-IC} case, a fast jet with Lorentz factor $\Gamma_{\rm j} = 300$ is launched with a delay of $t_{\rm delay} = 30$ sec, owing to episodic central engine  activity {\citep[e.g.,][]{ramirez-ruiz01a,ramirez-ruiz01b}}. 
(ii) In the {\it Late-IC} case, a slower jet with Lorentz factor $\Gamma_{\rm j} = 50$ is launched  with a delay of $t_{\rm delay} = 100$ sec, owing to late time central engine activity causing X-ray flares.
The cocoon physical properties as well as the thermal radiation produced are only weakly dependent on the jet Lorentz factor. Thus, the approach used to compute the IC emission is also nearly independent on the prompt jet Lorentz factor, which, for numerical limitations, is smaller in our simulations than in typical GRBs.

In both cases, we assume that the central engine stays active at a low luminosity level so as to keep the jet funnel open, so the delayed jet does not interact with the cocoon hydrodynamically.
If the shocked cocoon gas refills the jet funnel, then the delayed jet will be significantly decelerated if $\Gamma_j^2 (M_c c^2/\Gamma_c) > E_j$, where  $\Gamma_c$ is the cocoon Lorentz factor, $M_c$ is the cocoon mass within the jet funnel, and $E_j$ is the kinetic energy of the delayed jet. Modeling the possible hydrodynamical interaction between delayed jets and the cocoon is out of the scope of the current paper. Shocks at the jet-cocoon interface may accelerate relativistic electrons, which will inverse-Compton scatter the cocoon radiation and produce additional non-thermal spectrum. These effects need to be studied in detail in a future work.

In this section, all quantities in the comoving frame of the fluid cell are denoted with a prime and unprimed quantities are measured in the lab frame (rest frame of the central engine).

\subsection{Method}
For each snapshot from the simulation, we first identify the
position of the cocoon photosphere (equation  \ref{eq:1}).
Each photospheric cell has its own diffusive flux $F_{\rm ph}^\prime$ (given by flux-limited diffusion in equation \ref{flux}), temperature
$T_{\rm ph}^\prime$, Lorentz factor  $\Gamma$, and the angle between the fluid cell's velocity vector and the jet axis is $\arctan(v_r/v_z)$.  The intensity in the comoving frame of each photospheric cell is assumed  to be isotropic (equation
\ref{eq:intensity}) and the lab-frame intensity in an arbitrary direction
is given by the corresponding Lorentz transformation.

Then, for each scattering cell located in the hypothetic jet
at time $t$ at the position $\vec{R} (r, z)=(r,0,z)$,
we consider light rays coming from different directions
denoted by the unit vector $\vec{e}(\theta,\phi)$, where $\theta$ is
the polar angle and $\phi$ is the azimuth angle.
At an earlier time $t_{\rm 0}$ the position of the light ray is
\begin{equation}
  \label{eq:3}
  \vec{R}_{\rm 0}(t_{\rm 0}) = \vec{R} (r, z) - c(t - t_{\rm 0})
  \vec{e}(\theta,\phi).
\end{equation}
For each of the previous snapshots at $t_{\rm 0} <
t$, we do not expect $\vec{R}_{\rm 0}(t_{\rm 0})$ to be exactly on the
photosphere surface (since the snapshot times $\{t_{\rm 0}\}$ are
discrete), so the closest photospheric cell is
considered as where the light ray was emitted. Therefore, the
intensity of the cocoon's radiation field at
any time $t$ at the position of the scattering cell $\vec{R} (r, z)$ is given by
\begin{equation}
  \label{eq:13}
  I_{\nu_{\rm 0}} (\theta, \phi) = \epsilon  B_{\nu_{\rm 0}}\left(T = D_{\psi}T_{\rm
    ph}^\prime\right)
\end{equation}
where $T_{\rm ph}^\prime$ is the temperature of the photospheric cell found by the ray-tracing method above, $\epsilon = F_{\rm ph}^\prime/[\sigma_{\rm SB}(T^\prime)^4]$, $D_{\psi} =[\Gamma_{\rm 0} (1 - \beta_{\rm 0} \cos \psi)]^{-1}$ is the Doppler factor, and $\psi$ is the angle between $\vec{e}(\theta,\phi)$ and the velocity vector of the photospheric cell. Then, it is straightforward to integrate light rays of different directions and frequencies  to calculate the IC lightcurve and spectrum.

In the following, we describe the detailed integrating procedure presented in \citet{lu15}. Experienced readers may jump to the results in the next  subsection.

Consider a scattering cell at time $t$ and position $\vec{R} (r, z)$
with a total number of electrons denoted by $\d N_{\rm e}$. The cell
is moving at an angle $\alpha$ from the $z$-axis with velocity $\vec{v}$
which corresponds to a Lorentz factor $\Gamma = [1 -
(v/c)^2]^{-1/2}$. We define another set of Cartesian coordinates
$\tilde{x}\tilde{y}\tilde{z}$ in which $\hat{\tilde{z}}$ is
parallel to the cell's velocity vector $\vec{v}$, $\hat{\tilde{y}}$ is
parallel to the original $y$-axis $\hat{y}$,
and $\hat{\tilde{x}}$ is in the original $xz$ plane (which goes through the
scattering cell). The
relation between the two coordinate systems is described by a rotation
along the $y$-axis by an angle $\alpha$ according to the right-hand
rule, and the transformation matrix (in Cartesian coordinates) is
\begin{equation}
  \label{eq:5}
    \Lambda(\alpha) =
      \begin{pmatrix}
        \cos \alpha & 0 &\sin\alpha \\
         0 &1 &0\\
         -\sin \alpha &0 & \cos \alpha
      \end{pmatrix}.
\end{equation}
In this new coordinate system, the original direction of the light ray
$\vec{e}(\theta,\phi)$ becomes $\vec{\tilde{e}}(\thet, \phit) =
\Lambda \vec{e}$. This relation gives the mapping between
$(\theta,\phi)$ and $(\thet,\phit)$ as follows
\begin{equation}
  \label{eq:6}
  \begin{cases}
    \cos \theta & = \sin \thet \cos \phit \sin \alpha + \cos\thet \\
    \tan \phi &= \frac{\sin \thet \sin \phit}{\sin \thet \cos \phit
      \cos \alpha - \cos \thet \sin \alpha}
  \end{cases}
\end{equation}
Then, we convert the intensity of external radiation field
$I_{\nu_{\rm 0}} (\thet, \phit)$ to the comoving frame
$\tilde{x}^\prime \tilde{y}^\prime \tilde{z}^\prime$ of the scattering
cell by using the Lorentz transformations
\begin{equation}
  \label{eq:7}
  \begin{cases}
      & \nu_{\rm 0} = \Gamma(1 + \beta \cos \thetp) \nu_{\rm 0}^\prime =
  D\nu_0^\prime \\
  & \cos \thet = (\cos \thetp + \beta)/(1 + \beta \cos\thetp) \\
  & \phit = \phitp \\
  & I_{\nu_{\rm 0}^\prime}^\prime (\thetp, \phitp) = I_{\nu_{\rm 0}} (\thet, \phit)/D^3
  \end{cases}
\end{equation}
where we have defined a new Doppler factor $D$ in the first expression.
We are interested in the photons scattered towards the observer, who
in the original $xyz$ frame is located on the $z$-axis and in the
direction $(\theta_{\rm obs}^\prime, \phi_{\rm obs}^\prime)$ in the comoving
$\tilde{x}^\prime \tilde{y}^\prime \tilde{z}^\prime$ frame given by
\begin{equation}
  \label{eq:8}
  \begin{cases}
   & \cos \theta_{\rm obs}^\prime = (\cos \alpha - \beta)/(1 - \beta \cos
   \alpha) \\
   & \phi_{\rm obs}^\prime = \pi
  \end{cases}
\end{equation}
The number of photons scattered into the solid
angle $\Delta \Omega_{\rm obs}^\prime$ around the $(\theta_{\rm
  obs}^\prime, \phi_{\rm obs}^\prime)$ direction in a duration
$\mathrm{d} t^\prime$ and frequency range $\mathrm{d}\nu^\prime$ can
be obtained by integrating the radiation incoming at different
frequencies $\nu_{\rm 0}^\prime$ and from different directions
$\tilde{\Omega}^\prime = (\thetp, \phitp)$, i.e.
\begin{equation}
  \label{eq:9}
  \begin{split}
      \mathrm{d} N_\gamma =\ &\mathrm{d} N_{\rm e}  \mathrm{d} \nu^\prime
  \Delta \Omega_{\rm obs}^\prime \mathrm{d} t^\prime\\
  &\int \mathrm{d}
  \tilde{\Omega}^\prime  \int \mathrm{d} \nu_{\rm 0}^\prime
  \frac{I_{\nu_{\rm 0}^\prime}^\prime (\thetp, \phitp)}{h \nu_{\rm
      0}^\prime} \frac{\partial^2 \sigma}{\partial \nu^\prime \partial
  \Omega_{\rm obs}^\prime} \mathrm{e}^{-\tau},
  \end{split}
\end{equation}
where $\frac{\partial^2 \sigma}{\partial \nu^\prime \partial
  \Omega^\prime}$ is the differential cross section and the optical
depth $\tau$ describes the attenuation by the part of the jet that lies along the incident photons' trajectory before they enter the scattering cell.
These $\mathrm{d} N_\gamma$  photons will arrive at the observer at time
\begin{equation}
  \label{eq:10}
  t_{\rm obs} = t - z/c,
\end{equation}
and the lab-frame frequency is given by
\begin{equation}
  \label{eq:11}
  \nu =[\Gamma (1 - \beta \cos \alpha)]^{-1}\nu^\prime = D_{\alpha} \nu^\prime\;,
\end{equation}
where we have defined another Doppler factor $D_{\alpha}$.
Differentiating equation (\ref{eq:10}) gives $\d t_{\rm obs} =
(1 - \beta \cos \alpha) \mathrm{d} t =\d t^\prime
/D_{\alpha}$ and the Lorentz transformation of solid angles gives
$\Delta \Omega_{\rm obs}^\prime = D_{\alpha}^2 \Delta \Omega_{\rm obs}$
, so this scattering cell contributes a specific luminosity of
\begin{equation}
  \label{eq:12}
  \begin{split}
      \mathrm{d} L_\nu^{\rm iso} =\ &\mathrm{d} N_{\rm e}
  4\pi D_{\alpha}^2 h\nu \\
  & \int \mathrm{d}
  \tilde{\Omega}^\prime  \int \mathrm{d} \nu_{\rm 0}^\prime
  \frac{I_{\nu_{\rm 0}^\prime}^\prime (\thetp, \phitp)}{h \nu_{\rm
      0}^\prime} \frac{\partial^2 \sigma}{\partial \nu^\prime \partial
  \Omega_{\rm obs}^\prime} \mathrm{e}^{-\tau}
  \end{split}
\end{equation}
where we have used $\Delta \Omega_{\rm obs} = 4\pi$ for
isotropic equivalent luminosity. As integrating equation (\ref{eq:12})
is computationally expensive, we made the following two simplifications.
(i) When electrons are cold ($\gamma_{\rm e}^\prime = 1$), the
differential cross section is assumed to be isotropic and every
electron has Thomson cross section.
(ii) When electrons are hot ($\gamma_{\rm e}^\prime \gg 1$), we use
the full angle-dependent Klein-Nishina differential cross section, but
instead of the blackbody seed spectrum $I_{\nu_{\rm 0}^\prime}^\prime
(\thetp, \phitp) = B_{\nu_{\rm 0}^\prime}(D_{\psi}T_{\rm ph}^\prime/D)$,
we use a $\delta$-function centred at $h\nu_{\rm 0}^\prime = 2.7
D_{\psi}kT_{\rm ph}^\prime/D$.

At last, we add up the IC luminosity
from all the scattering cells in different snapshots weighted by their
individual numerical timesteps.

\subsection{Results}
In this subsection, we present the lightcurves and spectra for the Early-IC and Late-IC cases. In the Early-IC case, the jet is launched  at lab-frame time $t_{\rm  delay}=30\rm\ s$ with isotropic equivalent power $L_{\rm j}^{\rm iso} = 10^{52}\rm\ erg\ s^{-1}$ and Lorentz factor $\Gamma_{\rm j} = 300$. We have two sub-cases corresponding to $\gamma_e  = 1$ (cold electrons) and $\gamma_e = 10$ (hot electrons). In the Late-IC case, the jet is launched at $t_{\rm  delay}=100\rm\ s$ with Lorentz factor $\Gamma_{\rm j} = 50$, and we consider two luminosities $L_{\rm j}^{\rm iso} =  10^{51}$ and $10^{52}\rm\ erg\ s^{-1}$. The duration of the delayed jets in all cases is $t_{\rm j} = 10\rm\ s$,
which is motivated by the observation that the typical ratio between the duration and peak time for X-ray flares is 0.1{--}0.3 (e.g. Chincarini, Moretti, et al., 2007, ApJ, 671, 1903)

The calculations in all cases are done by using an opening angle of $2/\Gamma_j$. The flux contributed by high-latitude ($> 1/\Gamma_j$) regions is strongly suppressed due to relativistic beaming. We have also tested other opening angles of $1.5/\Gamma_j$ and $3/\Gamma_j$ and the differences are negligible.

We ignore the hydrodynamical interaction between the jet and cocoon and only consider IC scattering after the jet emerges from below the photosphere.

%%%%%%%%%%%%%%%%%%%%%%%%%%%%%%%%%%%%
\begin{figure}
\centering
\includegraphics[width=8cm]{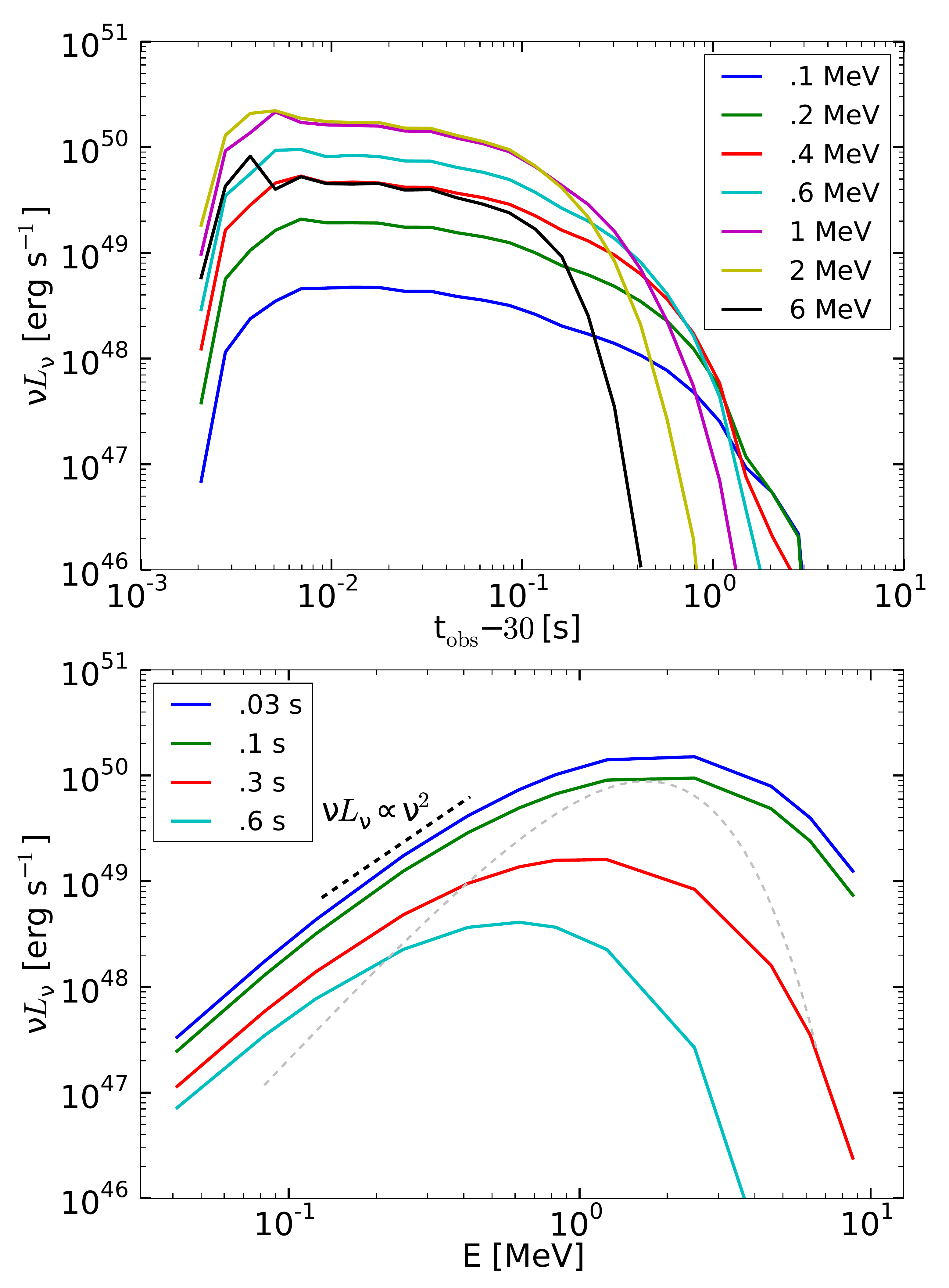}
\caption{Lightcurves and spectra of the Early-IC case with cold electrons, corresponding to the hydro-simulation with progenitor mass $5.45M_{\rm \odot}$ and prompt jet Lorentz factor 10. The hypothetical jet, launched with a delay of 30 s with respect to the prompt jet that produced the cocoon, has Lorentz  factor $300$ and isotropic kinetic power $10^{52}\rm\ erg\ s^{-1}$, duration $10\rm\ s$ (which does not really have an effect on the IC luminosity), and electrons in the jet are assumed to be cold ($\gamma_{\rm  e}=1$). We find that the IC luminosity is initially $\sim 10^{50}$-$10^{51}\rm\ erg\ s^{-1}$ when the jet emerges from the cocoon surface, but it then drops quickly as the jet surpasses the cocoon. The legend in the lower panel is $t_{\rm obs} - 30 \rm\ s$. The IC emission peaks at $\sim 3\rm\  MeV$ initially and then the peak frequency drops quickly with time because cocoon photons are moving increasingly parallel with the jet. The spectra are broader than blackbody (grey dashed line), because the low-frequency power-law is characteristic IC spectrum $\nu L_{\rm \nu}\propto \nu^2$ and the high-frequency part is broadened by multi-colour seed photons.}
\label{fig11}
\end{figure}
%%%%%%%%%%%%%%%%%%%%%%%%%%%%%%%%%%%%

The jet emerges from the cocoon surface at radius $R_{\rm em} = 10^{13} R_{\rm em,13}$ and $10^{13.5} R_{\rm em,13.5}\rm\ cm$ for the Early-IC and Late-IC cases respectively, so the peak time is
\begin{equation}
  \label{eq:14}
  \begin{split}
      t_{\rm peak} \simeq R_{\rm em}/(\Gamma_{\rm j}^2c)
 &\simeq 3.7\times10^{-3} R_{\rm em,13}  \Gamma_{\rm j, 2.5}^{-2} \rm\ s\\
&\simeq 0.4 R_{\rm em,13.5}  \Gamma_{\rm j, 1.7}^{-2} \rm\ s.
  \end{split}
\end{equation}
The thickness of the transparent shell at the jet front that external
photons can penetrate through is
\begin{equation}
  \label{eq:4}
  \begin{split}
      \Delta R_{\rm tr} = \frac{1}{n_{\rm e}^\prime \Gamma_{\rm j}
    \sigma_{\rm T}} &\simeq 7.7\times10^9 \frac{R_{13}^2\Gamma_{\rm
      j, 2.5}}{L_{\rm j, 52}^{\rm iso}} \rm\ cm \\
  &\simeq 1.1\times10^{11} \frac{R_{13.5}^2\Gamma_{\rm
      j, 1.7}}{L_{\rm j, 51}^{\rm iso}} \rm\ cm,
  \end{split}
\end{equation}
which needs to be compared to the size of the causally connected region given by
\begin{equation}
  \label{eq:15}
  \begin{split}
    R/\Gamma_{\rm j}^2 &\simeq 1.1\times10^{8} R_{13}\Gamma_{\rm j,2.5}^{-2}\ \mathrm{cm}\\
    &\simeq 1.2\times10^{10} R_{13.5}\Gamma_{\rm j,1.7}^{-2} \rm\ cm.
  \end{split}
\end{equation}
The duration of the peak IC luminosity is given by
\begin{equation}
  \label{eq:16}
  \Delta t_{\rm peak} \sim \frac{\mathrm{max}(\Delta R_{\rm tr},
    R/\Gamma_{\rm j}^2)}{c}.
\end{equation}
In the Early-IC case, we have $\Delta t_{\rm peak}\simeq 0.26\rm\ s$. In the Late-IC cases with $L_{\rm j}^{\rm iso} =10^{51}$ and $10^{52}\rm\ erg\ s^{-1}$, we have $\Delta t_{\rm peak}\sim 3.7$ and $0.4\rm\ s$ respectively. We plot the lightcurves and spectra in Figure \ref{fig11}-\ref{fig14} and the main results are summarized as follows (see the captions for details):

%%%%%%%%%%%%%%%%%%%%%%%%%%%%%%%%%%%%
\begin{figure}
\centering
\includegraphics[width=8cm]{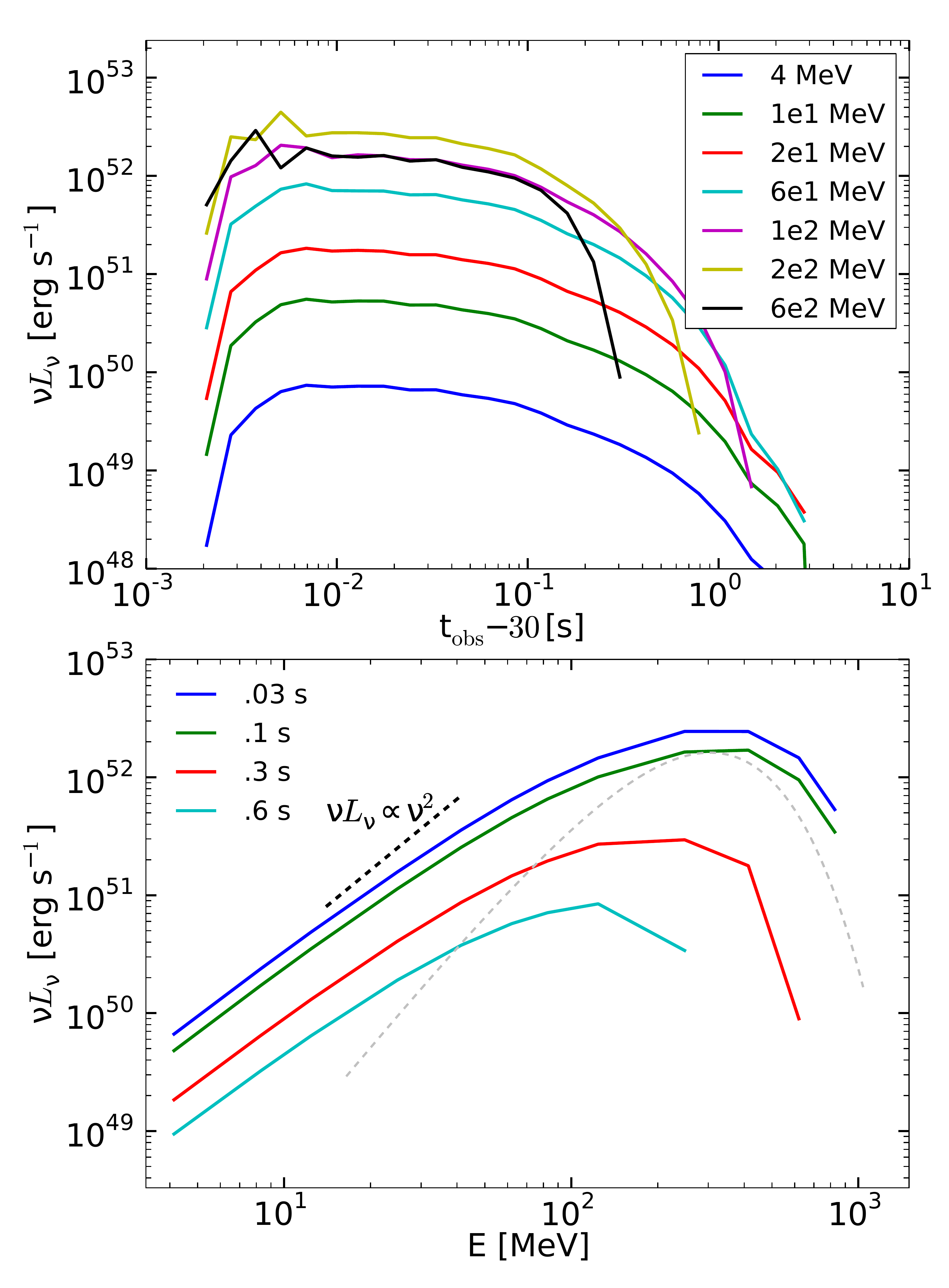}
\caption{Lightcurves and spectra of the Early-IC case with hot electrons, corresponding to the same hydro-simulation as in Figure (\ref{fig11}). The jet parameters are the same as in Figure (\ref{fig11}), but electrons are assumed to be monoenergetic with Lorentz factor $\gamma_{\rm e}=10$ in the jet comoving frame. The IC luminosity is a factor of $\sim\gamma_{\rm  e}^2$ higher than in the cold electron case. The IC luminosity exceeds the jet kinetic power for the first $0.3\rm\ sec$, which means the front of the jet must be Compton dragged to a lower Lorentz factor.  The legend in the lower panel is $t_{\rm obs} - 30 \rm\ s$. The peak of the IC spectra is at
$\sim 300\rm\ MeV$ initially, so we may expect some pair production from $\gamma\gamma$ interaction (which is not included in our calculation). The spectra are broader than blackbody (gray dashed line), with the low-frequency power-law being $\nu L_{\rm \nu}\propto \nu^2$. The relatively sharp drop off (compared to the cold-electron case in Figure \ref{fig11}) at the high-energy end is caused by our $\delta$-function approximation of the blackbody spectrum from each of the cocoon photospheric cells and the true spectrum should be broader.}
\label{fig12}
\end{figure}
%%%%%%%%%%%%%%%%%%%%%%%%%%%%%%%%%%%%

%%%%%%%%%%%%%%%%%%%%%%%%%%%%%%%%%%%%
\begin{figure}
\centering
\includegraphics[width=8cm]{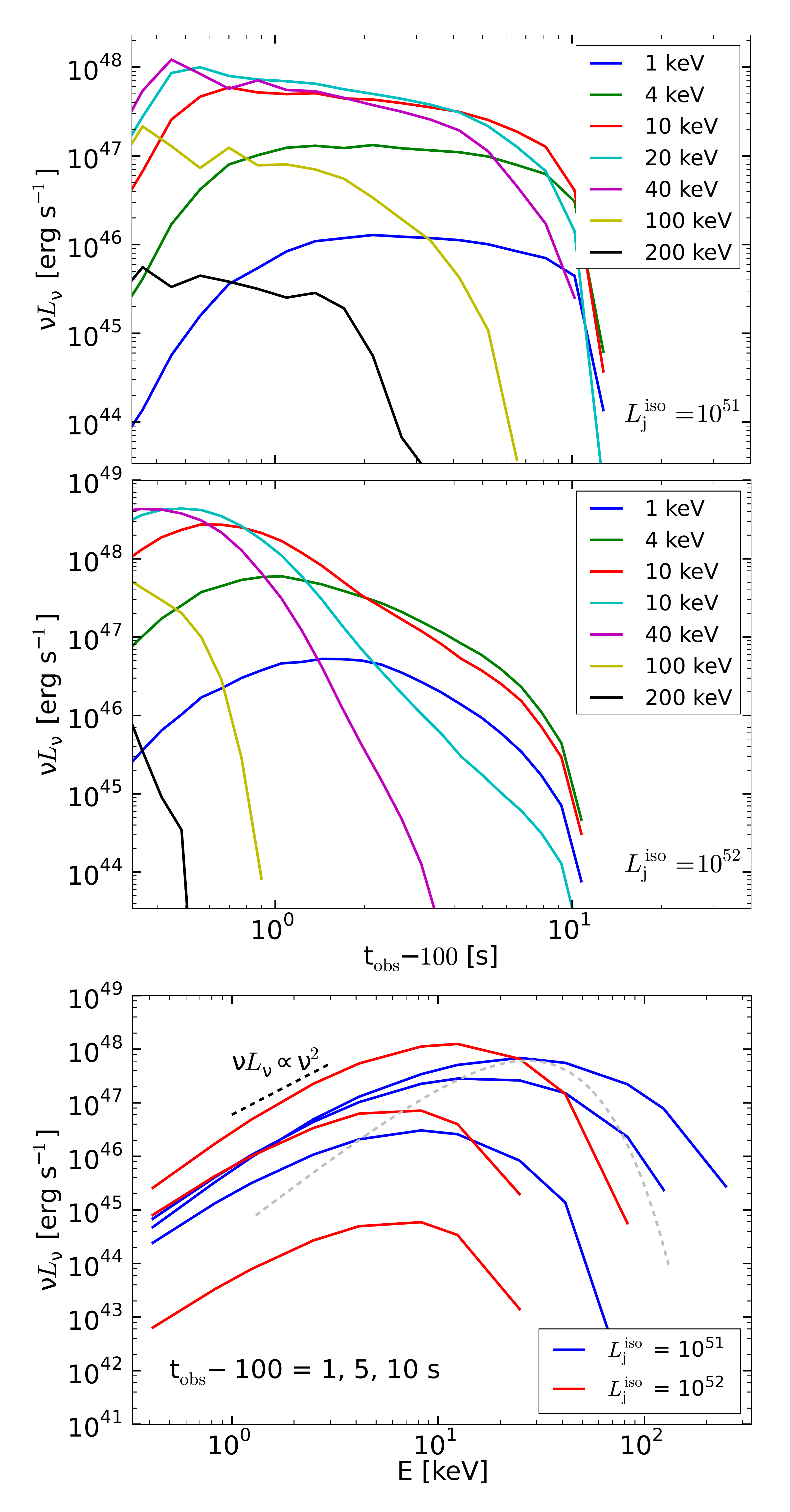}
\caption{Late-IC case with different isotropic jet power, corresponding to the hydro-simulation with progenitor mass $5.45M_{\rm \odot}$ and prompt jet Lorentz factor 10. The first two panels are lightcurves for $L_{\rm j}^{\rm iso} = 10^{51}$ (upper panel) and $10^{52}\rm\ erg\ s^{-1}$ (middle panel). Both jets have Lorentz factor $\Gamma_{\rm j}=50$, delay time $t_{\rm delay} = 100\rm\ s$ and duration $t_{\rm j} = 10\rm\ s$, and electrons are cold ($\gamma_{\rm e}=1$). In the upper panel, the IC luminosity stays nearly flat at $\sim10^{48}\rm\  erg\ s^{-1}$ for $\sim5$ seconds because the whole jet is scattering cocoon photons ($\Delta R_{\rm tr}\simeq 0.37 ct_{\rm j}$); but in the middle panel, only a small part of the jet is optically thin ($\Delta R_{\rm tr}\simeq 3.7\times10^{-2} ct_{\rm j}$), so the peak emission lasts shorter. The two cases have the same total fluence because the total number of scattered photons is the same. The sharp drop off at $>10\rm\ s$ is due to emission from large polar angle regions (``curvature effect''). The lower panel shows the spectra of the two cases evaluated at three different observer's time $t_{\rm obs} - 100= 1$, $5$ and $10\rm\ s$ (in order of decreasing luminosity). The spectra are broader than blackbody (gray dashed line), with the low-frequency power-law being $\nu L_{\rm \nu}\propto \nu^2$.
}
\label{fig13}
\end{figure}
%%%%%%%%%%%%%%%%%%%%%%%%%%%%%%%%%%%%

%%%%%%%%%%%%%%%%%%%%%%%%%%%%%%%%%%%%
\begin{figure}
\centering
\includegraphics[width=8cm]{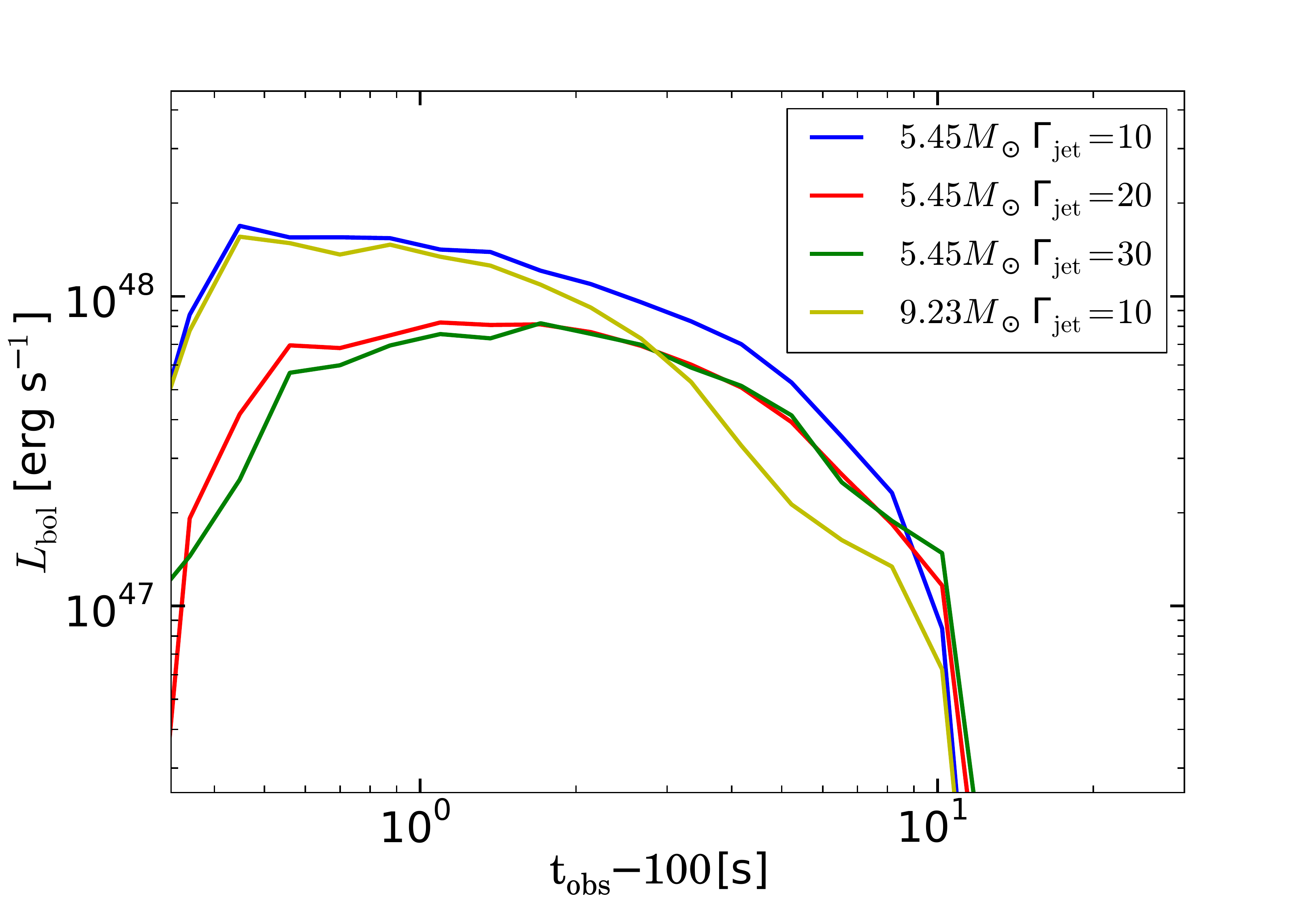}
\caption{Late-IC bolometric lightcurves for different hydro-simulations (as shown in the legend). In all four cases, the hypothetic delayed jet has isotropic power $L_{\rm j}^{\rm iso} = 10^{51}\rm\ erg\ s^{-1}$, Lorentz factor $\Gamma_{\rm  j}=50$, delay time $t_{\rm delay} = 100\rm\ s$ and duration $t_{\rm j} = 10\rm\ s$, and electrons are cold ($\gamma_{\rm e}=1$). The cocoon is moving at an increasingly larger Lorentz factor as the prompt jet Lorentz factor increases, and this decreases the IC luminosity. The progenitor's mass profile affects the time evolution of the IC emission, as seen in the difference between the blue and yellow line. The IC emission has a stronger dependence on the prompt jet Lorentz factor than on progenitor star density profile, so it may be challenging to infer progenitor proterties from the IC emission.
}
\label{fig14}
\end{figure}
%%%%%%%%%%%%%%%%%%%%%%%%%%%%%%%%%%%%

(1) When electrons are cold, IC scattering generally taps a fraction of a few$\times10^{-4}$ to $10^{-3}$ of the total energy of the delayed jet. This means the IC emission is likely overwhelmed by other radiation mechanisms
(it is not responsible for generating the majority of the emission observed in X-ray flares.)
 However, when there is a modest amount of jet dissipation so that the root mean squared Lorentz factor $\sqrt{<\gamma_{\rm e}^2>} \gtrsim 10$ at radii between $10^{13}$ and $10^{14}\rm\ cm$, the peak IC luminosity exceeds the jet luminosity. This means that a fraction of a few to 10 percent of the jet is strongly dragged by the IC force to a lower Lorentz factor. Further jet dissipation such as internal shocks or magnetic reconnection could be triggered by this IC drag. A significant fraction of the 0.1-1 GeV emission during the prompt phase may be contributed by IC emission off cocoon photons.

(2) The spectrum is always broader than blackbody, even when electrons are in monoenergetic distribution (cold or hot). A power-law electron distribution will lead to a power-law spectrum in the high frequency part. Therefore, the IC component off cocoon emission might have been missed in previous studies looking for a thermal component.

(3) The effect of progenitor star's mass and density profile has a
much weaker effect on the IC emission than the Lorentz factor of the prompt jet. This is due to the strong dependence of IC emission
on the cocoon Lorentz factor \citep{kumar14}. It may be challenging to infer progenitor properties from the IC emission. Since the IC emission is very sensitive to the Lorentz factor of the delayed jet (peak energy $\propto \Gamma_{\rm j}^2$ and luminosity $\propto \Gamma_{\rm j}^4$), IC flux could provide an independent measurement of the $\Gamma_{\rm j}$.

Finally, we notice that internal shocks moving close to the head of the jet should produce an IC emission similar to that obtained in the case of a 30 s delay. 

\section{Conclusions}
\label{sec:conclusions}

We have carried out numerical simulations of a long-GRB jet propagating
through the progenitor star and its wind, the production of a cocoon
that results from this interaction, and the spectrum and lightcurve of
the emergent cocoon radiation. The dynamical evolution of the jet/cocoon
was followed from $10^8$ cm to $\sim 3\times 10^{14}$ cm. We considered
two different progenitor stars of mass 5.45 M$_\odot$ and 9.23 M$_\odot$
(right before the collapse) with radius of $3\times10^{10}$ and $10^{11}$ cm
respectively; they were models E25 \citep{heger00} and 12TH \citep{woosley06b}.
The simulations were run for a luminosity $L_{\rm jet}$ = $2\times10^{50}$ erg s$^{-1}$
and several different jet Lorentz factors ($\Gamma=10$, 20, 30). In each of these cases
the jet duration was taken  to be 20 s.

The cocoon emission was calculated by post-processing results of the
numerical simulations. The cocoon spectrum is quasi-thermal and peaks in the X-ray band at $\sim 5$ keV ($\sim 0.5$ keV) a few seconds ($\sim 100$ s) after the cocoon emerges above the stellar surface. The bolometric luminosity of cocoon emission is $\sim 10^{47}$ erg s$^{-1}$ for about 200 s in the host galaxy rest frame ($\sim 10$ min in the observer frame for a typical redshift of 2); this luminosity is comparable to the GRB X-ray afterglow luminosity during the plateau phase which is observed in a good  fraction of long-GRBs starting at about 100 s after the prompt $\gamma$-ray emission ends. 

The cocoon lightcurve is nearly independent of the jet Lorentz factor when $\Gamma_{\rm jet} > 20$, but depends on the stellar structure (Figure \ref{fig6}). When a more extended stellar model is considered, the X-ray light curve increases at later times and is dimmer, while the spectra is softer at all observing times (see Figure \ref{fig9}).
We note that the velocity, density and energy distributions depend strongly on the jet initial conditions, in particular on the jet luminosity history (for instance, a jet with a larger luminosity can deposit larger amount of energies in the cocoon making it brighter in X-rays), on the presence of a magnetic field (which can provide extra collimation to the jet and thereby reduce the cocoon energy), on the stellar structure (as shown in this paper) and on the jet structure. In addition, the presence of large asymmetries in the jet (which need to be studied by three dimensional simulations and could be due, e.g., to precession or wiggling of the jet) would also affect the results by reducing the velocity of the cocoon then making the cocoon dimmer. Given the uncertainties in the jet characteristics, nevertheless, we showed in this paper that the thermal emission from a GRB cocoon could be detectable at least in some GRBs.

Thus, detection of a quasi-thermal component in the X-ray afterglow lightcurves of long-GRBs can be used, combined with constraints from optical observations from the associated jet-driven supernova and more detailed hydrodynamic calculations (e.g., including a more complete treament of the radiation transfer and a broad range of progenitor properties), to infer the density structure and radius of the progenitor star.

Observations of X-ray flares are usually intepreted as evidence of late energy injection.
A small fraction of photons from the cocoon pass through the relativistic jet and are inverse-Compton scattered by electrons in the jet to higher
energies. 
We assume that electrons are in monoenergetic distribution with Lorentz factor $\gamma_{\rm e}$ in the jet comoving frame. 
We computed the IC spectrum generated by the interaction between the cocoon photons and shocks with a delay of 30 and 100 s.
For a jet of Lorentz factor 300 launched with a delay time of 30 s, the resulting IC luminosity is of $\sim 3 \gamma_{\rm e}^2\%$ of jet luminosity and the IC spectrum peaks at $\sim 3\gamma_{\rm e}^2$ MeV when the jet emerges above the cocoon photosphere. Photons from the cocoon are also scattered by jets with longer delays (those responsible for
X-ray flares during the GRB afterglow phase), and the resulting IC luminosity
is $\sim0.3 \gamma_{\rm e}^2\%$ of late jet luminosity with the spectral peak at $\sim 10\gamma_{e}^2$ keV, if the late jet has a Lorentz
factor of 50. These results on IC scatterings of cocoon photons
by relativistic jets are consistent with the analytical calculations
of \citet{kumar14} and would provide an indirect measurement of the GRB jet Lorentz factor and magnetization parameter if detected.

%%%%%%%%%%%%%%%%%%%%%%%%%%%%%%%%%%%%%%%%%%%%%%%%%%%%%%%%%%%%%%%%%%%%%%%%%%%%%%
% ACKNOWLEDGMENTS
%%%%%%%%%%%%%%%%%%%%%%%%%%%%%%%%%%%%%%%%%%%%%%%%%%%%%%%%%%%%%%%%%%%%%%%%%%%%%%

\section*{Acknowledgements}

We thank Diego Lopez C\'amara and Hiro Nagataki for useful discussions.
F.D.C. thank the UNAM-PAPIIT grants IA103315 and IN117917. W.L. is funded by a graduate fellowship  (``Named Continuing Fellowship'') at the University of Texas at Austin. ER-R acknowledges financial support from the David and Lucile Packard Foundation and UCMEXUS (CN-12-578). The simulations were performed on the Miztli supercomputer at UNAM.

%%%%%%%%%%%%%%%%%%%%%%%%%%%%%%%%%%%%%%%%%%%%%%%%%%%%%%%%%%%%%%%%%%%%%%%%%%%%%%
% BIBLIOGRAPHY
%%%%%%%%%%%%%%%%%%%%%%%%%%%%%%%%%%%%%%%%%%%%%%%%%%%%%%%%%%%%%%%%%%%%%%%%%%%%%%

% Don't change these lines
\bsp	% typesetting comment
\label{lastpage}
\end{document}